\documentclass[11pt]{article}

\usepackage{amsmath,amssymb}
\usepackage{cite}
\usepackage[dvips]{color,graphicx}
\begin{document}
\title{\Large\bf Interplay between kaon condensation and  hyperons in highly dense matter}
\author{Takumi Muto\thanks{Email address: takumi.muto@it-chiba.ac.jp}\\
Department of Physics, Chiba Institute of Technology \\
        2-1-1 Shibazono, Narashino, Chiba 275-0023, Japan}

\date{\today}
\maketitle

\vspace{1.0cm}
      
\begin{abstract}
Possible coexistence and/or competition of kaon condensation with hyperons are investigated in hyperonic matter, where hyperons are mixed in the ground state of neutron-star matter. The formulation is based on the effective chiral Lagrangian for the kaon-baryon interaction and the nonrelativistic baryon-baryon interaction model. 
First, the onset condition of the $s$-wave kaon condensation realized from hyperonic matter is reexamined. It is shown that the usual assumption of the continuous phase transition is not always kept valid in the presence of the negatively charged hyperons ($\Sigma^-$). 
Second, the equation of state (EOS) of the kaon-condensed phase in hyperonic matter is discussed. In the case of the stronger kaon-baryon attractive interaction, it is shown that a local energy minimum with respect to the baryon number density appears as a result of considerable softening of the EOS due to both kaon condensation and hyperon-mixing and recovering of the stiffness of the EOS at very high densities. This result implies a possible existence of self-bound objects with kaon condensates on any scale from an atomic nucleus to a neutron star. 
\end{abstract}

\begin{description} 
{\footnotesize\item PACS: 05.30.Jp, 13.75.Jz, 26.60.+c, 95.35.+d 
\item Keywords:  kaon-baryon interaction, kaon condensation, hyperon, neutron stars, density isomer}
\end{description}

\newpage
\section{Introduction}
\label{sec:intro}

\ \ It has long been suggested that antikaon ($K^-$) condensation should be realized in high density hadronic matter\cite{kn86,mtt93,t95,lbm95,fmmt96,l96,pbpelk97}.\footnote{We consider antikaon ($K^-$) condensation, while we conventionally call it ``kaon condensation''.} It is characterized as a macroscopic appearance of strangeness in a strongly interacting kaon-baryon system, where chiral symmetry and its spontaneous breaking has a key role.
If kaon condensation exists in neutron stars, it softens the hadronic equation of state (EOS), having an influence on the bulk structure of neutron stars such as mass-radius relations\cite{fmmt96,l96,pbpelk97,tpl94}. Effects of the phase-equilibrium condition associated with the first-order phase transition 
on the inner structure of neutron stars have also been elucidated\cite{g01,hps93,cgs00,nr01,vyt03,mtv05}. With regard to dynamical evolution of newly-born neutron stars, delayed collapse of protoneutron stars accompanying a phase transition to kaon-condensed phase has been discussed\cite{bb94,bst95,p00,ty99}. The existence of kaon condensation is important for thermal evolution of neutron stars since the neutrino emission processes are largely enhanced in the presence of kaon condensates\cite{bk88,t88,pb90,fmtt94}. 

In the kaon-condensed phase in neutron stars, 
the net (negative) strangeness gets abundant as a consequence of chemical equilibrium for weak interaction processes, $n\rightleftharpoons p K^-$, $e^-\rightleftharpoons K^-\nu_e$. 
At threshold, the onset condition for kaon condensation has been given by\footnote{Throughout this paper, the units $\hbar$ = $c$ = 1 are used.}
\begin{equation}
\omega(\rho_{\rm B})=\mu\ , 
\label{eq:onset}
\end{equation}
where $ \omega(\rho_B)$ is the lowest $K^-$ energy obtained at the baryon number density $\rho_{\rm B}$ 
from the zero point of the $K^-$ inverse propagator, $D_K^{-1}(\omega; 
\rho_B)=0$, and $\mu$ is the charge chemical potential which is equal to both the antikaon chemical potential 
$\mu_K$ and electron chemical potential $\mu_e$ under the chemical equilibrium condition for the weak processes\cite{mt92,bkrt92}. 
This onset condition (\ref{eq:onset}) is based on the assumption of 
{\it continuous phase transition} : 
Kaon condensation sets in with zero amplitude at a  critical density, above which kaon condensates develop smoothly with increase in baryon number density $\rho_{\rm B}$.
It has been shown that the onset condition (\ref{eq:onset}) holds true even if hyperon ($Y$) particle-nucleon ($N$) hole excitation through the $p$-wave kaon-baryon interaction is taken into account\cite{m93,kvk95} in the ordinary neutron-star matter where only nucleons and leptons are in $\beta$ equilibrium.  

Concerning another hadronic phase including strangeness, hyperons ($\Lambda$, $\Sigma^-$, $\Xi^-$, $\cdots$) as well as nucleons and leptons have been expected to be mixed in the ground state of 
neutron-star matter \cite{g85,ekp95,sm96,phz99,h00,s00,h98,bbs98,v00,bg97,y02,t04}. We call the hyperon-mixed neutron-star matter {\it hyperonic matter} throughout this paper. 
 With regard to coexistence or competition of kaon condensation with hyperons in neutrons stars, it has been pointed out that onset density of the $s$-wave kaon condensation subject to the condition (\ref{eq:onset}) in hyperonic matter is shifted to a higher density\cite{pbpelk97,ekp95,sm96} : The electron  carrying the negative charge is replaced by the negatively charged hyperon, so that the charge chemical potential $\mu$ (=$\mu_e=(3\pi^2\rho_e)^{1/3}$) is diminished as compared with the case of neutron-star matter without hyperons. As a result, the lowest $K^-$ energy $\omega(\rho_{\rm B})$ meets the charge chemical potential $\mu$ at a higher density. Subsequently, several works on the onset density and the EOS for the kaon-condensed phase in hyperonic matter have been elaborated with the relativistic mean-field theory\cite{pbg00,bb01}, the quark-meson coupling models\cite{mpp05,hrh06}. Recently the in-medium kaon dynamics and mechanisms of kaon condensation stemming from the $s$ and $p$-wave kaon-baryon interactions in hyperonic matter have been investigated\cite{m02,kv03}. 
 
 It is emphasized here that most of the results on the onset mechanisms of kaon condensation in {\it hyperonic matter} rely on the same assumption as the case of the usual neutron-star matter where hyperons are not mixed, i.e., the assumption of the {\it continuous phase transition} with the help of Eq.~(\ref{eq:onset}). 
In this paper, we reexamine the onset condition of kaon condensation realized from hyperonic matter.  We consider the $s$-wave kaon condensation and incorporate the kaon-baryon interaction within the effective chiral Lagrangian.  The nonrelativistic effective baryon-baryon interaction is taken into account, and the parameters are determined so as to reproduce the nuclear saturation properties and baryon potential depths deduced from the recent hypernuclear experiments\cite{g04}. We demonstrate that the assumption of continuous phase transition cannot always be applied to the case where the {\it negatively charged hyperons} ($\Sigma^-$) are already present in the ground state of hyperonic matter, as a result of competition between the negatively charged kaons and hyperons.  It will be shown that, in the vicinity of the baryon density a little lower than that where Eq.~(\ref{eq:onset}) is satisfied, there exists already another energy solution, for which kaons are condensed without mixing of the $\Sigma^-$ hyperons, in addition to the usual energy solution corresponding to the noncondensed state with the $\Sigma^-$ mixing. 
In particular, in the case of the stronger kaon-baryon attractive interaction, there is a discontinuous transition between these two states in a small density interval.  
Thus, from a theoretical viewpoint, one ought to be careful about the previous results concerning coexistence and/or competition of kaon condensates and $\Sigma^-$ hyperons, although quantitative effects resulting from the discontinuous phase transition are small.\footnote{Our discussion is concentrated on obtaining the energy solutions in the presence of hyperons under the constraints relevant to neutron stars. We don't discuss here the prescription of the Gibbs condition for the phase equilibrium associated with a first-order 
phase transition\cite{g01,hps93,cgs00,nr01,vyt03,mtv05}. 
This issue will be reported elsewhere\cite{m06}. A first-order phase transition to the $K^-$-condensed phase has also been discussed in another context in Refs.~\cite{kvk95,kv03}.} 

The interplay between $K^-$ condensates and $\Sigma^-$ hyperons can also be revealed in the EOS and characteristic features of the fully-developed kaon-condensed phase such as density-dependence of particle fractions. In the case of the stronger kaon-baryon attractive interaction, we will see that there appears a local energy minimum with respect to the baryon number density (a density isomer state) as a consequence of considerable softening of the EOS due to both kaon condensation and hyperon-mixing and recovering of the stiffness of the EOS at very high densities due to the increase in the repulsive interaction between baryons. 
 
The paper is organized as follows. In Sec.~\ref{sec:form},  the formulation to obtain the effective energy density of the kaon-condensed phase in hyperonic matter is presented. In Sec.~\ref{sec:fraction}, numerical results for the composition of the noncondensed phase of hyperonic matter are given for the subsequent discussions. Section~\ref{sec:validity} is devoted to the discussion on the validity of the continuous phase transition. The results for the EOS of the kaon-condensed phase are given in Sec.~\ref{sec:eos}. In Sec.~\ref{sec:summary}, summary and 
concluding remarks are addressed. In the 
Appendix, we remark that the two sets of parameters used in this paper for the baryon-baryon interaction models give different behaviors for the onset of $\Lambda$ and $\Sigma^-$ hyperons in ordinary neutron-star matter.  

\section{Formulation}
\label{sec:form}

\subsection{Outline of the kaon-condensed matter}
\label{subsec:outline}

\ \ In order to simplify and to make clear the discussion about the interrelations between kaon condensation and hyperons, we consider the s-wave kaon condensation by putting the kaon momentum $|{\bf k}|$=0, and we also 
take into account only the proton ($p$), $\Lambda$, neutron ($n$), and $\Sigma^-$ of the octet baryons and the ultrarelativistic electrons 
for kaon-condensed hyperonic matter in neutron stars. 

Within chiral symmetry, the classical kaon field as an order parameter of the $s$-wave kaon condensation is chosen to be a spatially uniform type: 
\begin{equation}
\langle K^-\rangle=\frac{f}{\sqrt{2}}\theta e^{-i\mu_K t} \ , 
\label{eq:kaon-field}
\end{equation} 
where $\theta$ is the chiral angle as an amplitude of  condensation, and $f$($\sim f_\pi$=93 MeV) is the meson decay constant. 

We impose the charge neutrality condition and baryon number conservation, and construct the effective Hamiltonian density by introducing the charge chemical potential $\mu$ and the baryon number chemical potential $\nu$, respectively, as the Lagrange multipliers corresponding to these two conditions. 
The resulting effective energy density is then written in the form 
\begin{equation}
{\cal E}_{\rm eff}={\cal E}+\mu (\rho_p-\rho_{\Sigma^-}-\rho_{K^-}-\rho_e)+\nu(\rho_p+ \rho_\Lambda+\rho_n+\rho_{\Sigma^-}) \ , 
\label{eq:eff}
\end{equation}
where ${\cal E}$ is the total energy density of the kaon-condensed phase, and $\rho_i$ ($i$= $p$, $\Lambda$, $n$, $\Sigma^-$, $K^-$, $e^-$) are the number densities of the particles $i$. It is to be noted that the number density of the kaon condensates $\rho_{K^-}$ consists of the free kaon part and the kaon-baryon interaction part of the vector type.[See Eq.~(\ref{eq:rhok}).]
From the extremum conditions for ${\cal E}_{\rm eff}$ with respect to variation of $\rho_i$, one obtains the following relations, 
\begin{subequations}\label{eq:chemeq}
\begin{eqnarray}
\mu_K&=&\mu_e=\mu_n-\mu_p=\mu_{\Sigma^-}-\mu_n =\mu \ ,\label{eq:chemeq1} \\
\mu_\Lambda&=&\mu_n=-\nu \ , \label{eq:chemeq2}
\end{eqnarray}
\end{subequations}
where $\mu_i$ ($i$= $p$, $\Lambda$, $n$, $\Sigma^-$, $K^-$, $e^-$) are the chemical potentials, which are given by $\mu_i=\partial{\cal E}/\partial\rho_i$. 
Equations~(\ref{eq:chemeq1}) and (\ref{eq:chemeq2}) imply that the system is in chemical equilibrium for the weak interaction processes, $n\rightleftharpoons pK^-$, $n\rightleftharpoons pe^-(\bar\nu_e)$, $ne^-\rightleftharpoons \Sigma^-(\nu_e)$, and $n\rightleftharpoons \Lambda(\nu_e\bar\nu_e)$. 

\subsection{Kaon-baryon interaction}
\label{subsec:kbint}

\ \ We are based on chiral symmetry for kaon-baryon interaction and start with the effective chiral SU(3)$_L \times$ SU(3)$_R$   
Lagrangian\cite{kn86}.\footnote{Except for setting 
$|{\bf k}|$=0, the basic formulation presented here is the same as that in Ref.~\cite{m02}, where both $s$-wave and $p$-wave kaon-baryon interactions are incorporated.} Then the relevant Lagrangian density, leading to the total energy density ${\cal E}$, consists of the following parts:
\begin{eqnarray}
{\cal L}&=&\frac{1}{4}f^2 \ {\rm Tr} 
\partial^\mu\Sigma^\dagger\partial_\mu\Sigma 
+\frac{1}{2}f^2\Lambda_{\chi{\rm SB}}({\rm Tr}M(\Sigma-1)+{\rm h.c.}) 
\cr
&+&{\rm Tr}\overline{\Psi}(i{\not\partial}-M_{\rm B})\Psi+{\rm 
Tr}\overline{\Psi}i\gamma^\mu\lbrack V_\mu, \Psi\rbrack
\cr
&+&a_1{\rm Tr}\overline{\Psi}(\xi M^\dagger\xi+{\rm h.c.})\Psi 
+ a_2{\rm Tr}\overline{\Psi}\Psi(\xi M^\dagger\xi+{\rm h.c.})+
a_3({\rm Tr}M\Sigma +{\rm h.c.}){\rm Tr}\overline{\Psi}\Psi \ , 
\label{eq:lag}
\end{eqnarray}
where the first and second terms on the r.~h.~s. of Eq.~(\ref{eq:lag}) are the kinetic and mass terms of mesons, respectively. $\Sigma$ is the nonlinear meson field defined by $\Sigma\equiv e^{2i\Pi/f}$, where $\displaystyle\Pi\equiv\sum_{a=1\sim 8}\pi_aT_a$ with $\pi_a$ being the octet meson fields and $T_a$ being the SU(3) generators. 
Since only charged kaon condensation is considered, the $\Pi$ is simply given as 
\begin{eqnarray}
\Pi=\frac{1}{\sqrt{2}}\left(
\begin{array}{ccc}
0 & 0 & K^+ \\
0 & 0 & 0 \\
K^- & 0 & 0 \\
\end{array}\right) \ .
\label{eq:meson}
\end{eqnarray}
In the second term of Eq.~(\ref{eq:lag}), $\Lambda_{\chi{\rm SB}}$ is the chiral symmetry breaking scale, $\sim$ 1 GeV, $M$ the mass matrix which is given by 
$M\equiv {\rm diag}(m_u, m_d, m_s)$ with the quark masses $m_i$. 
The third term in Eq.~(\ref{eq:lag}) denotes the free baryon part, where the $\Psi$ is the octet baryon field including only the $p$, $\Lambda$, $n$, $\Sigma^-$,
and $M_{\rm B}$ the baryon mass generated as a consequence of spontaneous chiral symmetry breaking.
The fourth term in Eq.~(\ref{eq:lag}) gives the $s$-wave kaon-baryon interaction of the vector type corresponding to the Tomozawa-Weinberg term with $V_\mu$ being the mesonic vector current defined by $V_\mu\equiv 
1/2(\xi^\dagger\partial_\mu\xi+\xi\partial_\mu\xi^\dagger)$ with $\xi\equiv \Sigma^{1/2}$. 
 The last three terms in Eq.~(\ref{eq:lag}) give the 
$s$-wave meson-baryon interaction of the scalar type, which explicitly breaks chiral symmetry.\footnote{The same types of the scalar and vector interactions are derived in the quark-meson coupling model\cite{tstw98}.} 

The quark masses $m_i$ are chosen to be $m_u$ = 6 MeV, $m_d$ = 12 MeV, and $m_s$ = 240 MeV. Together with these values, the parameters $a_1$ and $a_2$ are fixed to be $a_1$ = $-$0.28, $a_2$ = 0.56 so as to reproduce the empirical octet baryon mass splittings\cite{kn86}. The parameter $a_3$ is related to the kaon-nucleon ($KN$) sigma terms simulating the $s$-wave $KN$ attraction of the scalar type through the expressions, $\Sigma_{Kp}=-(a_1+a_2+2a_3)(m_u+m_s)$, $\Sigma_{Kn}=-(a_2+2a_3)(m_u+m_s)$, evaluated at the on-shell Cheng-Dashen point for the effective chiral Lagrangian (\ref{eq:lag}). 
Recent lattice calculations suggest the value of the $KN$ sigma term $\Sigma_{KN}$=(300$-$400) MeV\cite{dll96}. We take the value of $a_3=-0.9$, leading to $\Sigma_{Kn}$=305 MeV, as a standard value. 
For comparison, we also take another value $a_3=-0.7$, which leads to $\Sigma_{Kn}$=207 MeV. 
The $K^-$ optical potential in symmetric nuclear matter, 
$V_{\rm opt}(\rho_{\rm B})$, is estimated as a scale of the $K^-$-nucleon attractive interaction. It is defined by
\begin{equation}
V_{\rm opt}(\rho_{\rm B})=\Pi_{K^-}(\omega(\rho_{\rm B}),\rho_{\rm B})
/2 \omega(\rho_{\rm B}) \ , 
\label{eq:vopt}
\end{equation}
where $\Pi_{K^-}\left(\omega(\rho_{\rm B}),\rho_{\rm B}\right)$
is the $K^-$ self-energy 
at given $\rho_B$ with $\rho_p=\rho_n=\rho_{\rm B}/2$. 
For $a_3=-0.9$ ($a_3=-0.7$), $V_{\rm opt}(\rho_0)$ is estimated to be $-$ 115 MeV ($-$ 95 MeV) at the nuclear saturation density $\rho_0$ (=0.16 fm$^{-3}$). 
 
In order to be consistent with the on-shell $s$-wave $K$ ($\bar K$)-$N$ scattering lengths, 
we have to take into account the range terms proportional to $\omega^2$ coming from the higher-order terms in chiral expansion and a pole contribution from the 
$\Lambda$(1405)\cite{lbm95,fmmt96}. Nevertheless, these contributions to the energy density become negligible in high-density matter. 
Therefore, we omit these correction terms throughout this paper and consider the simplified expression for the energy density of the kaon-condensed phase. 
 
\subsection{Effective energy density}
\label{subsec:energy}

\ \ The total effective energy density ${\cal E}_{\rm eff}$ is separated into baryon, meson, and lepton parts as
${\cal E}_{\rm eff}={\cal E}_{\rm eff}^{\rm B}+{\cal E}_{\rm eff}^{\rm M}+{\cal E}_{\rm eff}^{\rm e}$. 
The kaon-baryon interaction is incorporated in the baryon part ${\cal E}_{\rm eff}^{\rm B}$, and the meson part ${\cal E}_{\rm eff}^{\rm M}$ consists of the free classical kaons only. The baryon part ${\cal E}_{\rm eff}^{\rm B}$ and the meson part ${\cal E}_{\rm eff}^{\rm M}$ are derived from the effective chiral Lagrangian (\ref{eq:lag}). 
After the nonrelativistic reduction for the baryon part of the effective Hamiltonian by way of the Foldy-Wouthuysen-Tani transformation and with the mean-field approximation, one obtains
\begin{equation}
{\cal E}_{\rm eff}^{\rm B}=\sum_{i=p,\Lambda,n,\Sigma^-} \sum_{\stackrel{|{\bf p}| \leq 
|{\bf p}_F(i)|}{s=\pm1/2}}E_{{\rm eff},s}^{(i)} ({\bf p}) \ , 
\label{eq:be}
\end{equation}
where ${\bf p}_F(i)$ are the Fermi momenta, and the 
subscript `$s$' stands for the spin states for the baryon. 
The effective single-particle energies $E_{{\rm eff},s}^{(i)} ({\bf p})$ for the baryons $i$ are represented by
\begin{subequations}\label{eq:spe}
\begin{eqnarray} 
E_{{\rm eff},s}^{(p)} ({\bf p})&=& {\bf p}^2/2M_N -(\mu+\Sigma_{Kp})(1-\cos\theta)+\mu+\nu \ ,  \label{eq:spep} \\
E_{{\rm eff},s}^{(\Lambda)} ({\bf p})&=& {\bf p}^2/2M_N -\Sigma_{K\Lambda}(1-\cos\theta)
+\delta M_{\Lambda N}+\nu  \ , \label{eq:spel} \\
E_{{\rm eff},s}^{(n)} ({\bf p})&=&{\bf p}^2/2M_N-\Big(\frac{1}{2}\mu+\Sigma_{Kn}\Big)(1-\cos\theta)+\nu \ , \label{eq:spen} \\
E_{{\rm eff},s}^{(\Sigma^-)} ({\bf p})&=& {\bf p}^2/2M_N-\Big(-\frac{1}{2}\mu+\Sigma_{K\Sigma^-}\Big)(1-\cos\theta)
+\delta M_{\Sigma^- N}-\mu+\nu \ , \label{eq:spes} 
\end{eqnarray}
\end{subequations}
where $M_N$ is the nucleon mass, $\delta M_{\Lambda N}$ (= 176 MeV) is the $\Lambda$-$N$ mass difference and $\delta M_{\Sigma^- N}$ (= 258 MeV) the $\Sigma^-$-$N$ mass difference. The ``kaon-hyperon sigma terms'' are defined by $\displaystyle\Sigma_{K\Lambda}\equiv -\left(\frac{5}{6}a_1+\frac{5}{6}a_2+2a_3\right)(m_u+m_s)$ 
and $\displaystyle \Sigma_{K\Sigma^-}\equiv -(a_2+2a_3)(m_u+m_s)$ (=$\Sigma_{Kn}$). It is to be noted that each term in Eqs.~(\ref{eq:spe}) contains both the kaon-baryon attraction of the scalar type simulated by the ``sigma term'' and the kaon-baryon interaction of the vector type proportional to $\mu$ the coefficient of which is given by the V-spin charge of each baryon.  
 
The meson contribution to the effective energy density, ${\cal E}_{\rm eff}^{\rm M}$, is given by 
the substitution of the classical kaon field (\ref{eq:kaon-field}) 
into the meson part of the effective Hamiltonian :  
\begin{equation}
{\cal E}_{\rm eff}^{\rm M}=-\frac{1}{2}f^2\mu^2\sin^2\theta+f^2m_K^2(1-\cos\theta) \ , 
\label{eq:me}
\end{equation}
where $m_K\equiv [\Lambda_{\chi {\rm SB}}(m_u+m_s)]^{1/2}$, which is 
identified with the free kaon mass, and is replaced by the 
experimental value, 493.7 MeV. The lepton contribution to the effective energy density is given as
\begin{equation}
{\cal E}_{\rm eff}^{\rm e}
=\frac{\mu^4}{4\pi^2}-\mu\frac{\mu^3}{3\pi^2}
=-\frac{\mu^4}{12\pi^2}
\label{eq:ee}
\end{equation}
with $\rho_e=\mu^3/(3\pi^2)$ for the ultrarelativistic electrons. 

\subsection{Baryon potentials}
\label{subsec:pot}

\ \ We introduce a potential energy density ${\cal E}_{\rm pot}$ as a local effective baryon-baryon interaction, which is assumed to be given by functions of the number densities of the relevant baryons\cite{bg97}. 
In order to take into account the baryon potential effects on both the whole energy of the system and the baryon single-particle energies consistently, we take the following prescription: The baryon potential $V_i$ ($i=p, \Lambda$, $n$, $\Sigma^-$) is defined as 
\begin{equation}
V_i=\partial{\cal E}_{\rm pot}/\partial\rho_i 
\label{eq:vi}
\end{equation}
with $\rho_i$ being the number density of baryon $i$, and it is added to each effective single particle energy, $E_{{\rm eff},s}^{(i)}({\bf p})\rightarrow {E'}_{{\rm eff},s}^{(i)}({\bf p})=E_{{\rm eff},s}^{(i)}({\bf p})+V_i$. The potential energy density ${\cal E}_{\rm pot}$ is added to the total effective energy density ${\cal E}^{\rm eff}$, and the term $\displaystyle\sum_{i=p, \Lambda, n,\Sigma^-}\rho_i V_i$ is subtracted to avoid the double counting of the baryon interaction energies in the sum over the effective single particle energies ${E'}_{{\rm eff},s}^{(i)}({\bf p})$. 
Accordingly, the baryon part of the effective energy density is modified as 
\begin{eqnarray}
{\cal E}_{\rm eff}'^{\rm B}&=&\sum_{i=p, \Lambda, n, \Sigma^-} \sum_{\stackrel{|{\bf p}| \leq 
|{\bf p}_F(i)|}{s=\pm1/2}}{E'}_{{\rm eff},s}^{(i)} ({\bf p})+{\cal E}_{\rm 
pot}-\sum_{i=p, \Lambda, n, \Sigma^-} \rho_i V_i \cr
&=&\frac{3}{5}\frac{(3\pi^2)^{2/3}}{2M_N} (\rho_p^{5/3}+\rho_\Lambda^{5/3}+\rho_n^{5/3}+\rho_{\Sigma^-}^{5/3})+(\rho_\Lambda\delta M_{\Lambda p}+\rho_{\Sigma^-}\delta M_{\Sigma^- n}) + {\cal E}_{\rm pot}\cr
&-&\Bigg\lbrace\rho_p(\mu+\Sigma_{Kp})+\rho_\Lambda\Sigma_{K\Lambda}+\rho_n\Big(\frac{1}{2}\mu+\Sigma_{Kn}\Big)
+\rho_{\Sigma^-}\Big(-\frac{1}{2}\mu+\Sigma_{K\Sigma^-}\Big)\Bigg\rbrace(1-\cos\theta) \cr
&+& \mu(\rho_p-\rho_{\Sigma^-})+\nu\rho_{\rm B} \ .
\label{eq:eb}
\end{eqnarray}

The total effective energy density ${\cal E}'_{\rm eff}$ is obtained as the sum of the baryon, meson, and lepton parts whose explicit forms are given 
by Eqs.~(\ref{eq:eb}), (\ref{eq:me}), and (\ref{eq:ee}), respectively. 
For later convenience, we also show the total energy density ${\cal E}'$ including the potential contribution for baryons:
\begin{eqnarray}
 {\cal E}'&=&\frac{3}{5}\frac{(3\pi^2)^{2/3}}{2M_N}\Big(\rho_p^{5/3}+\rho_\Lambda^{5/3}+\rho_n^{5/3}+\rho_{\Sigma^-}^{5/3}\Big) \cr
 &+&(\rho_\Lambda\delta M_{\Lambda p}+\rho_{\Sigma^-}\delta M_{\Sigma^- n}) +{\cal E}_{\rm pot} \cr
 &-&\left(\rho_p \Sigma_{Kp}+\rho_\Lambda \Sigma_{K\Lambda}+\rho_n \Sigma_{Kn}+\rho_{\Sigma^-} \Sigma_{K\Sigma^-}\right)(1-\cos\theta) \cr
 &+&\frac{1}{2}f^2\mu^2\sin^2\theta+f^2m_K^2(1-\cos\theta)
 +\mu^4/(4\pi^2) \ , 
 \label{eq:te2}
\end{eqnarray}
where the first term on the right hand side denotes the baryon kinetic energy, the second term comes from the mass difference between the hyperons and nucleons, the third term the baryon potential energy, the fourth term the $s$-wave kaon-baryon scalar interaction brought about by the kaon-baryon sigma terms, the fifth and sixth terms the free parts of the condensed kaon energy (kinetic energy and free mass), and the last term stands for the lepton kinetic energy. 

For the hyperonic matter composed of $p$, $\Lambda$, $n$, and $\Sigma^-$, the potential energy density ${\cal E}_{\rm pot}$ is given by 
\begin{eqnarray}
{\cal E}_{\rm pot}&=&\frac{1}{2}\Big\lbrack a_{\rm NN}(\rho_{\rm 
p}+\rho_{\rm n})^2+b_{\rm NN}(\rho_{\rm p}-\rho_{\rm n})^2
+c_{\rm NN}(\rho_{\rm p}+\rho_{\rm n})^{\delta+1} \Big\rbrack\cr
&+& a_{\rm \Lambda N}(\rho_{\rm p}+\rho_{\rm n}){\rho_\Lambda}+c_{\rm 
\Lambda N}\Bigg\lbrack\frac{(\rho_{\rm p}
+\rho_{\rm n})^{\gamma+1}}{\rho_{\rm p}+\rho_{\rm 
n}+{\rho_\Lambda}}{\rho_\Lambda}
+\frac{{\rho_\Lambda}^{\gamma+1}}{\rho_{\rm p}
+\rho_{\rm n}+{\rho_\Lambda}}(\rho_{\rm p}
+\rho_{\rm n})\Bigg\rbrack + \frac{1}{2}( a_{YY}{\rho_\Lambda}^2
+c_{\rm YY}{\rho_\Lambda}^{\gamma+1}) \cr 
&+&a_{\rm \Sigma N}(\rho_{\rm p}+\rho_{\rm 
n}){\rho_{\Sigma^-}}+b_{\rm \Sigma N}(\rho_{\rm n}-\rho_{\rm 
p}){\rho_{\Sigma^-}} + c_{\rm \Sigma N}\Bigg\lbrack\frac{(\rho_{\rm p}
+\rho_{\rm n})^{\gamma+1}}{\rho_{\rm p}+\rho_{\rm 
n}+{\rho_{\Sigma^-}}}{\rho_{\Sigma^-}}
+\frac{{\rho_{\Sigma^-}}^{\gamma+1}}{\rho_{\rm p}
+\rho_{\rm n}+\rho_{\Sigma^-}}(\rho_{\rm p}
+\rho_{\rm n})\Bigg\rbrack \cr
&+&a_{\rm YY}{\rho_{\Sigma^-}}{\rho_\Lambda}+c_{\rm YY}\Bigg\lbrack 
\frac{{\rho_{\Sigma^-}}^{\gamma+1}}{{\rho_{\Sigma^-}}+{\rho_\Lambda}}{\rho_\Lambda}+\frac{{\rho_\Lambda}^{\gamma+1}}{{\rho_{\Sigma^-}}+{\rho_\Lambda}}{\rho_{\Sigma^-}}\Bigg\rbrack 
+\frac{1}{2}\Big\lbrack (a_{\rm YY}
+b_{\Sigma\Sigma}){\rho_{\Sigma^-}}^2
+c_{\rm YY}{\rho_{\Sigma^-}}^{\gamma+1}\Big\rbrack \ .
\label{eq:epot}
\end{eqnarray}
 The parameters in the potential energy density (\ref{eq:epot}) are determined as follows: (i) The parameters $a_{NN}$ and $c_{NN}$ in the $NN$ part are fixed so as to reproduce the standard nuclear saturation density $\rho_0$=0.16 fm$^{-3}$ and the binding energy $-$16 MeV in symmetric nuclear matter. With the parameters $a_{NN}$, $c_{NN}$, and $\delta$, the incompressibility $K$ in symmetric nuclear matter is obtained. The parameter $b_{NN}$ for the isospin-dependent term in the $NN$ part is chosen to reproduce the empirical value of the symmetry energy $\sim$ 30 MeV at $\rho_B=\rho_0$. (ii) For the $YN$ parts, $a_{\Lambda N}$ and $c_{\Lambda N}$ are basically taken to be the same as those in Ref.~\cite{bg97}, where the single $\Lambda$ orbitals in ordinary hypernuclei are reasonably fitted. The depth of the $\Lambda$ potential in nuclear matter is then given as $V_{\Lambda}(\rho_p=\rho_n=\rho_0/2)=a_{\Lambda N}\rho_0+c_{\Lambda N}\rho_0^\gamma$=$-$27 MeV\cite{mdg88}.  
The depth of the $\Sigma^-$ potential $V_{\Sigma^-}$ in nuclear matter is taken to be repulsive, following recent theoretical calculations\cite{kf00,fk01} and the phenomenological analyses on the ($K^-$, $\pi^\pm $) reactions at BNL\cite{b99,d99}, ($\pi^-$, $K^+$) reactions at KEK\cite{n02,dr04,hh05}, and the $\Sigma^-$ atom data\cite{mfgj95}: $V_{\Sigma^-}(\rho_p=\rho_n=\rho_0/2)=a_{\Sigma N}\rho_0+c_{\Sigma N}\rho_0^\gamma$=23.5 MeV and $b_{\Sigma N}\rho_0$=40.2 MeV. This choice of the parameters corresponds to the values in Ref.~\cite{d99} based on the Nijmegen model F. 
(iii) Since the experimental information on the $YY$ interactions is not enough, we take the same parameters for the $YY$ part as those in Ref.~\cite{bg97}. 

Taking into account the conditions (i) $\sim$ (iii), we adopt the following two parameter sets throughout this paper: (A) $\delta$=$\gamma$=5/3. In this case, one obtains $K$=306 MeV, which is larger than the standard empirical value 210$\pm$30 MeV\cite{b80}. 
(B) $\delta$=4/3 and $\gamma$=2.0. From the choice $\delta$=4/3, one obtains $K$=236 MeV which lies within the empirical value. The choice $\gamma$=2.0 leads to the stiffer EOS for hyperonic matter at high densities compared with the case (A). 
Numerical values of the parameter sets (A) and (B) are listed in Table~\ref{tab:para}. Here we abbreviate the EOS for hyperonic matter with the use of (A) and (B) as H-EOS (A) and H-EOS (B), respectively. 
\begin{table}[h]
\caption{Parameters in the potential energy density. ($^{\rm 
a}$MeV$\cdot$fm$^3$, \  $^{\rm b}$MeV$\cdot$fm$^{3\gamma}$, \  $^{\rm c}$MeV$\cdot$fm$^{3\delta}$)}
\label{tab:para}
\begin{center}
\begin{tabular}{c|cr|cr|cr}
\hline
H-EOS & parameter & & parameter & & parameter & \\ \hline
 & $\gamma$ & 5/3 &  ${a_{\Lambda N}}^{\rm a}$ & $-$387.0 & 
${a_{YY}}^{\rm a}$ & $-$552.6 \\
 & $\delta$ & 5/3 & ${c_{\Lambda N}}^{\rm b}$ & 738.8 & ${c_{YY}}^{\rm 
b}$ & 1055.4   \\
(A) & ${a_{NN}}^{\rm a}$ & $-$914.2 & ${a_{\Sigma N}}^{\rm a}$ & $-$70.9 &  
${b_{\Sigma\Sigma}}^{\rm a}$ & 428.4 \\
 & ${b_{NN}}^{\rm a}$ & 212.8 & ${b_{\Sigma N}}^{\rm a}$ & 251.3 & 
 &   \\
 & ${c_{NN}}^{\rm c}$ & 1486.4 & ${c_{\Sigma N}}^{\rm b}$ & 738.8 & 
 &  \\\hline
  & $\gamma$ & 2 &  ${a_{\Lambda N}}^{\rm a}$ & $-$342.8 & 
${a_{YY}}^{\rm a}$ & $-$486.2 \\
 & $\delta$ & 4/3 & ${c_{\Lambda N}}^{\rm b}$ & 1087.5 & ${c_{YY}}^{\rm 
b}$ & 1553.6   \\
(B) & ${a_{NN}}^{\rm a}$ & $-$1352.3 & ${a_{\Sigma N}}^{\rm a}$ & $-$27.1 &  
${b_{\Sigma\Sigma}}^{\rm a}$ & 428.4 \\
 & ${b_{NN}}^{\rm a}$ & 212.8 & ${b_{\Sigma N}}^{\rm a}$ & 251.3 & 
 &   \\
 & ${c_{NN}}^{\rm c}$ & 1613.9 & ${c_{\Sigma N}}^{\rm b}$ & 1087.5 & 
 &  \\\hline
\end{tabular}
\end{center}
\end{table}

\subsection{Physical constraints}
\label{subsec:conditions}

\ \ The energy density and physical quantities in the ground state are obtained variationally by the extremization of the total effective energy density ${\cal E}_{\rm eff}'$ with respect to $\theta$, $\mu$, and each number density of the baryon $i$ at a given density $\rho_{\rm B}$. From $\partial{\cal E}_{\rm eff}'/\partial\theta=0$, one obtains the classical field equation for $\theta$,  
\begin{equation}
\sin\theta\Bigg\lbrack \mu^2\cos\theta-m_K^2+\frac{\mu}{f^2}\Big(\rho_p+\frac{1}{2}\rho_n-\frac{1}{2}\rho_{\Sigma^-}\Big)+\frac{1}{f^2}\sum_{i=p, \Lambda, n, \Sigma^-}\rho_i\Sigma_{Ki}\Bigg\rbrack=0 \ . 
\label{eq:theta}
\end{equation}
From $\partial{\cal E}_{\rm eff}'/\partial\mu=0$, one obtains the charge neutrality condition, 
\begin{equation}
\rho_p-\rho_{\Sigma^-}-\rho_{K^-}-\rho_e=0 \ , 
\label{eq:charge}
\end{equation}
where the number density of the kaon condensates $\rho_{K^-}$ is given as 
\begin{equation}
\rho_{K^-} = \mu f^2\sin^2\theta+\left(\rho_p+\frac{1}{2}\rho_n-\frac{1}{2}\rho_{\Sigma^-}\right)(1-\cos\theta) \ .
\label{eq:rhok}
\end{equation}
From $\partial{\cal E}_{\rm eff}'/\partial\nu=\rho_{\rm B}$, one obtains the baryon number conservation, 
\begin{equation}
\sum_{i=p, \Lambda, n, \Sigma^-}\rho_i=\rho_{\rm B} \ .
\end{equation}
The chemical equilibrium conditions for the weak interaction processes (\ref{eq:chemeq}), $n\rightleftharpoons p e^-(\bar\nu_e$), 
$n\rightleftharpoons \Lambda(\nu_e\bar\nu_e)$, $ne^-\rightleftharpoons \Sigma^-(\nu_e)$, are rewritten as: 
\begin{subequations}\label{eq:wequil}
\begin{eqnarray}
\mu_n&=&\mu_p+\mu \ ,  \label{eq:wequil1} \\
\mu_\Lambda&=&\mu_n \ , \label{eq:wequil2} \\
\mu_{\Sigma^-}&=&\mu_n+\mu  \ , \label{eq:wequil3} 
\end{eqnarray} 
\end{subequations}
where the chemical potentials for the baryons are given by $\mu_i=\partial{\cal E}'/\partial\rho_i$ with the help of Eqs.~(\ref{eq:te2}), (\ref{eq:theta}) and (\ref{eq:rhok}) :
\begin{subequations}\label{eq:mu}
\begin{eqnarray} 
\mu_n& = &\frac{(3\pi^2\rho_n)^{2/3}}{2M_N} -\Big(\frac{1}{2}\mu+\Sigma_{Kn}\Big)(1-\cos\theta)+V_n \ ,  \label{eq:mun} \\
\mu_p& = & \frac{(3\pi^2\rho_p)^{2/3}}{2M_N} -(\mu+\Sigma_{Kp})(1-\cos\theta)+V_p \ , \label{eq:mup} \\
\mu_\Lambda& = & \frac{(3\pi^2\rho_\Lambda)^{2/3}}{2M_N} -\Sigma_{K\Lambda}(1-\cos\theta)
+\delta M_{\Lambda N}+V_\Lambda \ , \label{eq:mul} \\
\mu_{\Sigma^-}& = &  \frac{(3\pi^2\rho_{\Sigma^-})^{2/3}}{2M_N} -\Big(-\frac{1}{2}\mu+\Sigma_{K\Sigma^-}\Big)(1-\cos\theta)
+\delta M_{\Sigma^- N}+V_{\Sigma^-} \ . \label{eq:mus} 
\end{eqnarray}
\end{subequations}

\section{Composition of matter in the noncondensed phase}
\label{sec:fraction}

\ \ The critical density satisfying Eq.~(\ref{eq:onset}) depends sensitively on the density dependence of $\mu$, which is also affected by the matter composition through the relation $\mu=\mu_e=(3\pi^2 \rho_e)^{1/3}$. 
\begin{figure}[!]
\noindent\begin{minipage}[t]{0.50\textwidth}
\begin{center}
\includegraphics[height=.3\textheight]{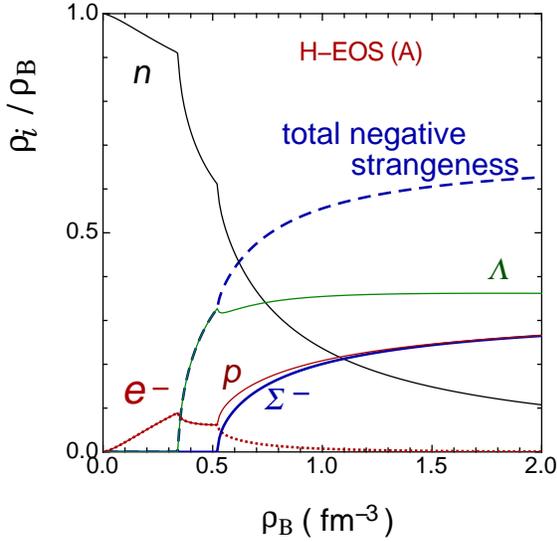}
\caption{Particle fractions $\rho_i/\rho_{\rm B}$ in the noncondensed hyperonic matter as functions of baryon number density $\rho_{\rm B}$. The H-EOS (A) is used. }
\label{fig:frac-ha}
\end{center}
\end{minipage}~
\begin{minipage}[t]{0.50\textwidth}
\begin{center}
\includegraphics[height=.3\textheight]{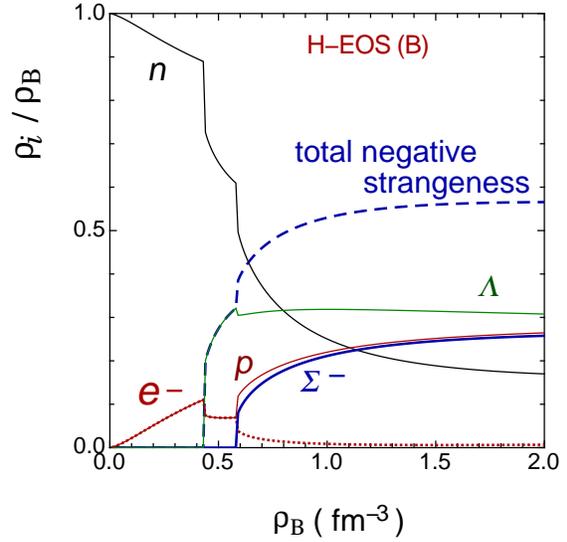}
\caption{The same as in Fig.~1, but for H-EOS (B). }
\label{fig:frac-hb}
\end{center}
\end{minipage}
\end{figure}
Thereby, before going into detail on the onset density of kaon condensation, we address behaviors of particle fractions in the noncondensed hyperonic matter. 
 
 In Figs.~\ref{fig:frac-ha} and \ref{fig:frac-hb}, particle fractions $\rho_i/\rho_{\rm B}$ ($i=p, \Lambda, n, \Sigma^- $, $e^-$) in the noncondensed hyperonic matter are shown as functions of baryon number density $\rho_{\rm B}$ for H-EOS (A) and (B), respectively. In both figures, the dashed lines stand for the ratio of the total negative strangeness number density $\rho_{\rm strange}$(=$\rho_\Lambda+\rho_{\Sigma^-}$) to the baryon number density $\rho_{\rm B}$.  

In the case of H-EOS (A), the $\Lambda$ hyperon starts to be mixed at $\rho_{\rm B}$ = $\rho_{\rm B}^c(\Lambda)$ = 0.340 fm$^{-3}$ (= 2.13 $\rho_0$) and the $\Sigma^-$ hyperon does at a higher density, $\rho_{\rm B}$= $\rho_{\rm B}^c(\Sigma^-)$ = 0.525 fm$^{-3}$ (= 3.28 $\rho_0$) (Fig.~\ref{fig:frac-ha}). In the case of H-EOS (B) , both hyperons start to be mixed at higher densities than the case of H-EOS (A), i.e., at $\rho_{\rm B}$ = $\rho_{\rm B}^c(\Lambda)$ $\sim$ 0.44 fm$^{-3}$ (= 2.69 $\rho_0$) for the $\Lambda$ and $\rho_{\rm B}$ = $\rho_{\rm B}^c(\Sigma^-)$ $\sim$ 0.59 fm$^{-3}$ (= 3.69 $\rho_0$) for the $\Sigma^-$, respectively (Fig.~\ref{fig:frac-hb}).   In fact, the condition for the 
$\Lambda$-mixing in the ($n$, $p$, $e^-$) matter, for instance, is written by the use of Eqs.~(\ref{eq:mun}) and (\ref{eq:mul}) as 
\begin{equation}
\delta M_{\Lambda N}+V_\Lambda \leq \frac{(3\pi^2\rho_n)^{2/3}}{2M_N}+V_n \ , 
\label{eq:lambda}
\end{equation}
where $V_\Lambda=a_{\Lambda N}\rho_{\rm B}+c_{\Lambda N}\rho_{\rm B}^\gamma$ and $\displaystyle V_n=a_{NN}\rho_{\rm B}-b_{NN}(\rho_p-\rho_n)+\frac{1}{2}c_{NN}(\delta+1)\rho_{\rm B}^\delta$. 
 In the case of H-EOS (B), the index $\delta$ (=4/3) is smaller than that for H-EOS (A) (=5/3), which makes the repulsive interaction of $V_n$ smaller than that for H-EOS (A). Furthermore the index $\gamma$ (=2) is bigger than that for H-EOS (A) (=5/3), which makes the repulsive interaction of $V_\Lambda$ larger than that for H-EOS (A). 
 Both effects push up the threshold density for the condition (\ref{eq:lambda}) as compared with the case of H-EOS (A). 
 
In general, the smaller value of the index $\delta$ 
simulating the higher order terms of the repulsive nucleon-nucleon interactions gives the smaller potential energy contributions for the nucleons. The larger value of the index $\gamma$ simulating the higher order repulsive terms of the hyperon-nucleon and hyperon-hyperon interactions gives the large potential energy contributions for the hyperons. As a result, the beta equilibrium conditions for the hyperons, $n\rightleftharpoons \Lambda \ (\nu_e\bar\nu_e)$, $n e^-\rightleftharpoons \Sigma^- \ (\nu_e)$, are satisfied at higher densities for H-EOS (B) than the case of H-EOS (A). 

It should be noted that, in the case of H-EOS (B), the $\Lambda$ and $\Sigma^-$ start to appear in the ground state of neutron-star matter such that the mixing ratios, $\rho_\Lambda/\rho_{\rm B}$, $\rho_{\Sigma^-}/\rho_{\rm B}$, increase discontinuously from zero to finite nonzero values above certain densities. This is a different behavior from the usual one, where the hyperon-mixing ratios increase continuously from zero as density increases like the case of H-EOS (A). 
(See the Appendix.) 

One can see common behavior with regard to the density dependence of each particle fraction for H-EOS (A) and H-EOS (B) : As the hyperons dominate the matter composition with increase in $\rho_{\rm B}$, the electron fraction decreases. In particular, the negative charge of the electron is taken over by that of the $\Sigma^-$ hyperon, so that the electron fraction decreases rapidly, while the $\Sigma^-$ fraction increases as the density increases.
 The proton fraction increases so as to compensate the negative charge of the $\Sigma^-$. At high densities, the $\Sigma^-$ and proton fractions amount to (20$-$30) \%,  the $\Lambda$ fraction to (30$-$40) \%, and the fraction of total negative strangeness to (50$-$60) \%. As a result of increase in fractions of the proton, $\Lambda$, and $\Sigma^-$, the neutron fraction decreases rapidly with increase in $\rho_{\rm B}$.
 
\section{Validity of continuous phase transition}
\label{sec:validity}

\ \ In ordinary neutron-star matter without hyperon-mixing, the onset density for kaon condensation is given by the condition, $\omega=\mu$ [Eq.~(\ref{eq:onset}) ]. Here the lowest energy $\omega$ for $K^-$ is obtained from the zero point of the inverse propagator for $K^-$, $D_K^{-1}(\omega; \rho_{\rm B})$, which can be read from expansion of the total effective energy density ${\cal E}_{\rm eff}'$ with respect to the chiral angle $\theta$ around $\theta=0$:
\begin{equation}
{\cal E}_{\rm eff}'(\theta)={\cal E}_{\rm eff}'(0)-\frac{f^2}{2}D_K^{-1}(\mu; \rho_{\rm B})\theta^2+O(\theta^4) \ .
\label{eq:dkinv}
\end{equation}
This onset condition, $D_K^{-1}(\mu; \rho_{\rm B})=0$, is equal to the nontrivial classical kaon-field equation (\ref{eq:theta}) with $\theta$ = 0, and is based on the assumption of the continuous phase transition : The chiral angle $\theta$, for instance, increases continuously  from zero as $\rho_{\rm B}$ increases. 
In this section, we consider validity of the assumption of the continuous phase transition to $K^-$ condensation in hyperonic matter.
Numerical results are presented by the use of the H-EOS (A) and H-EOS(B) for the noncondensed hyperonic matter EOS in Secs.~4.1 and 4.~2, respectively. 

\begin{figure}[!]
\begin{center}
\includegraphics[height=.3\textheight]{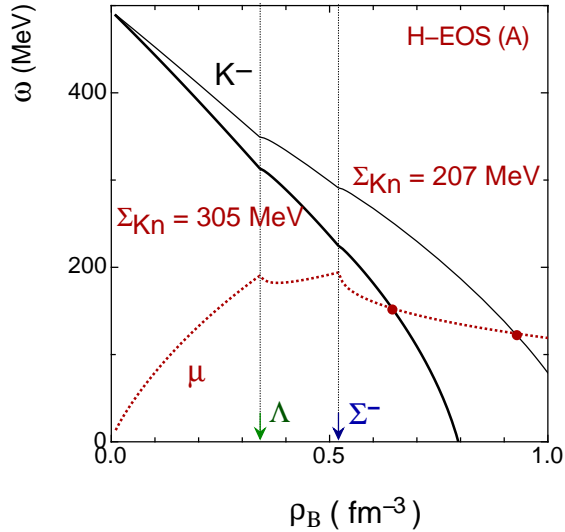}
\caption{The lowest energies of $K^-$ as functions of baryon number density $\rho_{\rm B}$ for $\Sigma_{Kn}$ = 305 MeV (bold solid line) and $\Sigma_{Kn}$ = 207 MeV (thin solid line) in the case of H-EOS (A). }
\label{fig:w-a}
\end{center}
\end{figure}

\subsection{Case of H-EOS (A) }
\label{subsec:a}

\ \ In Fig.~\ref{fig:w-a}, we show the lowest energies of the $K^-$ as functions of baryon number density $\rho_{\rm B}$ for $\Sigma_{Kn}$ = 305 MeV (bold solid line) and $\Sigma_{Kn}$ = 207 MeV (thin solid line) in the case of H-EOS (A). The dependence of the charge chemical potential $\mu$ (=$\mu_K=\mu_e$) on $\rho_{\rm B}$ is shown by the dotted line. The density at which the lowest $K^-$ energy $\omega$ crosses the $\mu$ is denoted as 
$\rho_{\rm B}^{c(2)}(K^-)$.
The charge chemical potential $\mu$ decreases with increase in density after the appearance of the negatively charged hyperon $\Sigma^-$, as seen in Fig.~\ref{fig:w-a}, so that the onset condition Eq.~(\ref{eq:onset}) is satisfied at a higher density than the case of neutron-star matter without mixing of hyperons.   From Fig.~\ref{fig:w-a}, one reads $\rho_{\rm B}^{c(2)}(K^-)$=0.6433~fm$^{-3}$ (=4.02$\rho_0$) for $\Sigma_{Kn}$=305 MeV and $\rho_{\rm B}^{c(2)}(K^-)$=0.9254~fm$^{-3}$ (=5.78$\rho_0$) for $\Sigma_{Kn}$=207 MeV. 

\begin{figure}[!]
\noindent\begin{minipage}[l]{0.50\textwidth}
\begin{center}
\includegraphics[height=.30\textheight]{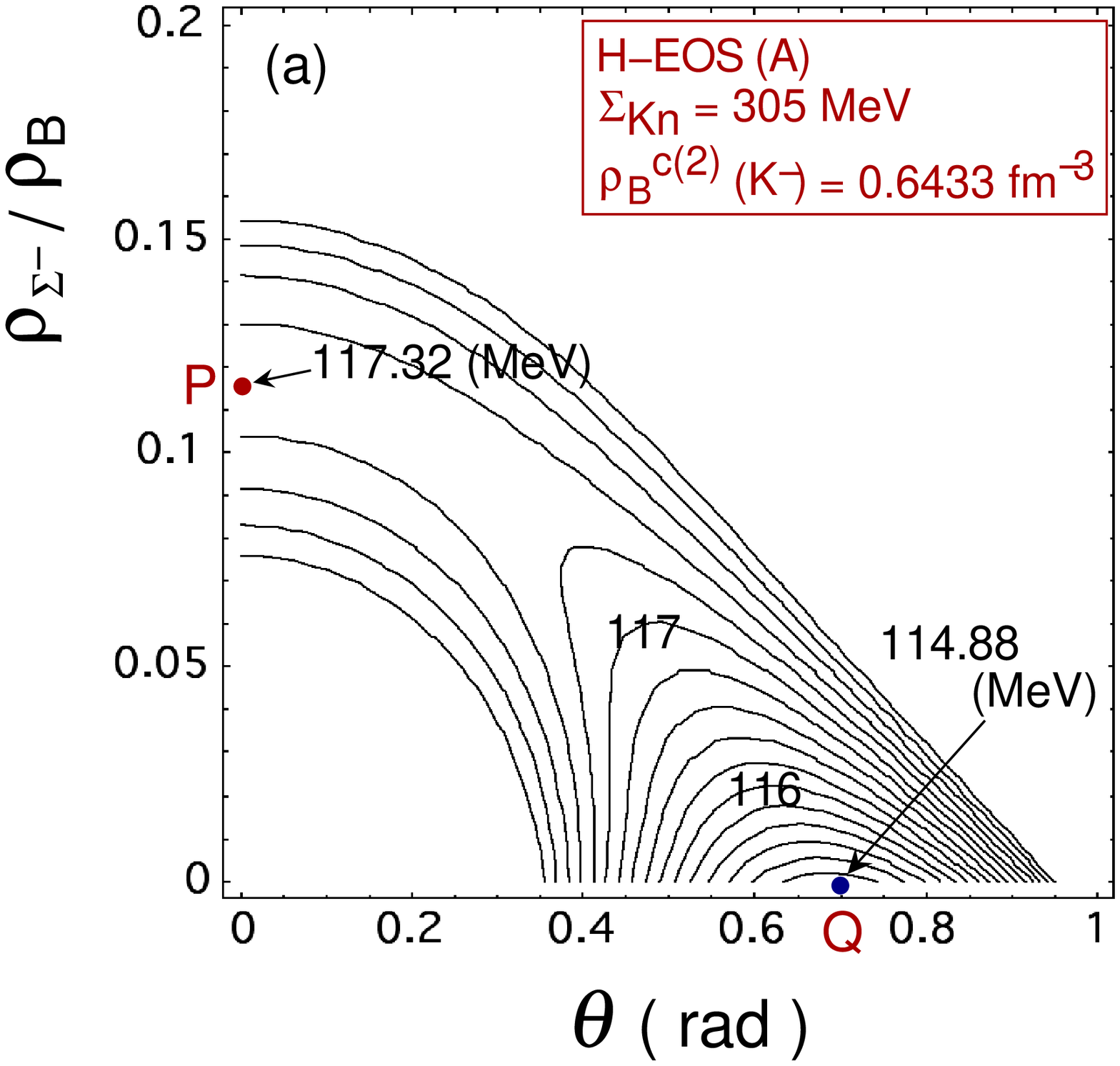}
\end{center}
\end{minipage}~
\begin{minipage}[r]{0.50\textwidth}
\begin{center}
\includegraphics[height=.30\textheight]{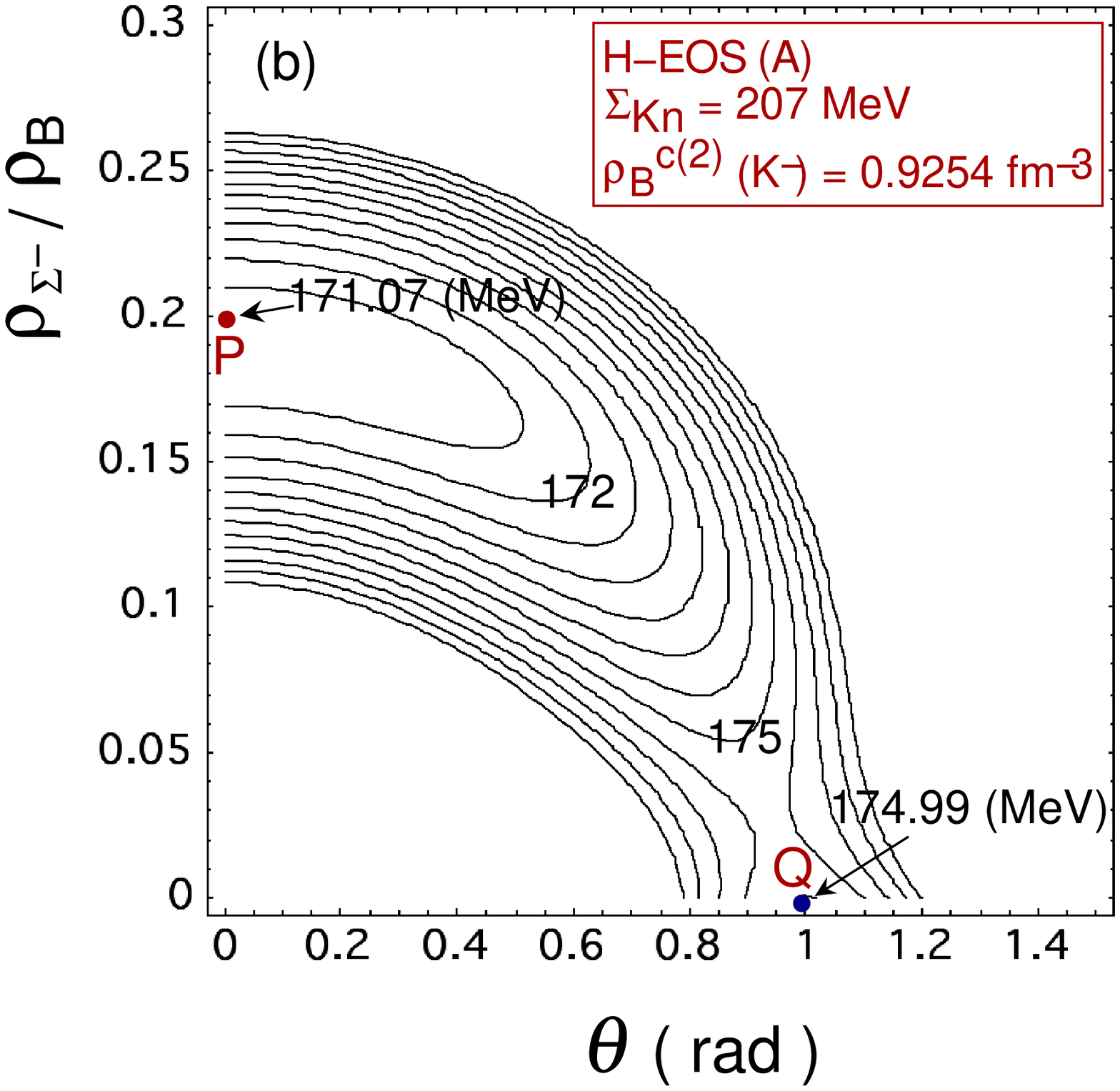}
\end{center}
\end{minipage}
\caption{(a) Contour plot of the total energy per baryon ${\cal E}'/\rho_{\rm B}$ in the ($\theta$, $\rho_{\Sigma^-}/\rho_B$) plane at $\rho_{\rm B}=\rho_{\rm B}^{c(2)}(K^-)$ for $\Sigma_{Kn}$=305 MeV in the case of H-EOS (A).  The energy interval is taken to be 0.2 MeV.  (b) The same as in (a), but for $\Sigma_{Kn}$=207 MeV. The energy interval is taken to be 0.5 MeV. See the text for details.}
\label{fig:contour-a}
\end{figure}

Now we examine whether the state at $\rho_{\rm B}=\rho_{\rm B}^{c(2)}(K^-)$ is the true ground state or not, by  considering the dependence of the total energy of the system at $\rho_{\rm B}=\rho_{\rm B}^{c(2)}(K^-)$ on the chiral angle $\theta$ and $\Sigma^-$-mixing ratio $\rho_{\Sigma^-}/\rho_B$ . In Fig.~\ref{fig:contour-a}, the 
contour plots of the total energy per baryon ${\cal E}'/\rho_{\rm B}$ 
in the ($\theta$, $\rho_{\Sigma^-}/\rho_B$) plane at $\rho_{\rm B}=\rho_{\rm B}^{c(2)}(K^-)$ are depicted for  
$\Sigma_{Kn}$=305 MeV [Fig.~\ref{fig:contour-a}(a)] and $\Sigma_{Kn}$=207 MeV [Fig.~\ref{fig:contour-a}(b)] in the case of H-EOS (A). Note that ${\cal E}'/\rho_{\rm B}$ has been maximized with respect to $\mu$ and minimized with respect to the other remaining parameters $\rho_\Lambda/\rho_{\rm B}$ and $\rho_p/\rho_{\rm B}$.  The energy interval between the contours is taken to be 0.2 MeV for $\Sigma_{Kn}$=305 MeV and 0.5 MeV 
for $\Sigma_{Kn}$=207 MeV. 
For $\Sigma_{Kn}$=305 MeV [Fig.~\ref{fig:contour-a}(a)], one obtains a state satisfying the condition $\omega$=$\mu$ at a point, ($\theta$, $\rho_{\Sigma^-}/\rho_B$)=(0, 0.117), where ${\cal E}'/\rho_{\rm B}$=117.32 MeV (denoted as P). However, this point is not a minimum, but a saddle point in the ($\theta$, $\rho_{\Sigma^-}/\rho_B$) plane.  A true minimum state exists at a different point, ($\theta$, $\rho_{\Sigma^-}/\rho_B$)=(0.70 rad, 0) in the plane (denoted as Q). This state Q stands for the fully-developed $K^-$-condensed state with no $\Sigma^-$-mixing. 
\begin{figure}[!]
\begin{center}
\includegraphics[height=.3\textheight]{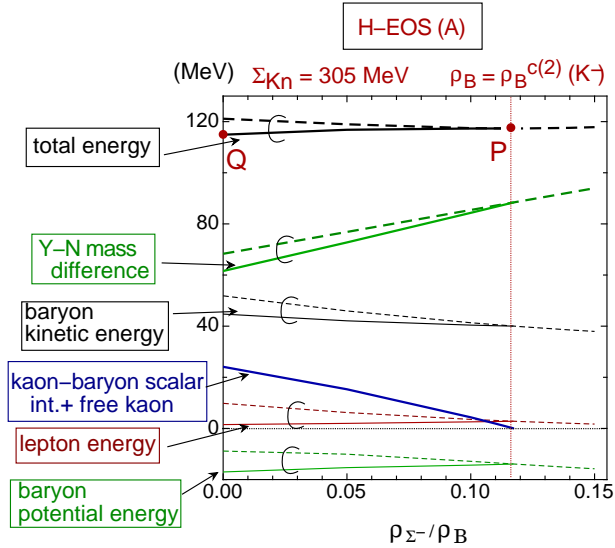}
\end{center}
\caption{The contributions from each term on the r.h.s. of Eq.~(12) to the total energy per baryon ${\cal E}'/\rho_{\rm B}$  as functions of the $\Sigma^-$-mixing ratio $\rho_{\Sigma^-}/\rho_{\rm B}$ at $\rho_{\rm B}$=$\rho_{\rm B}^{c(2)}(K^-)$ (=0.6433 fm$^{-3}$) for $\Sigma_{Kn}$=305 MeV (the solid lines). For comparison, those for the noncondensed state ($\theta$=0) is shown by the dashed lines. The H-EOS (A) is used for the hyperonic matter EOS. See the text for details.}
\label{fig:contrib-1}
\end{figure}

In Fig.~\ref{fig:contrib-1}, the total energy per baryon ${\cal E}'/\rho_{\rm B}$ and the contributions to ${\cal E}'/\rho_{\rm B}$ from each term on the r.h.s. of Eq.~(\ref{eq:te2}) are shown as functions of the $\Sigma^-$-mixing ratio $\rho_{\Sigma^-}/\rho_{\rm B}$ at $\rho_{\rm B}$=$\rho_{\rm B}^{c(2)}(K^-)$  (= 0.6433 fm$^{-3}$) for $\Sigma_{Kn}$=305 MeV by the solid lines. At a given $\rho_{\Sigma^-}/\rho_{\rm B}$ the total energy per baryon ${\cal E}'/\rho_{\rm B}$ is minimized with respect to $\theta$, $\rho_p$, $\rho_\Lambda$, and maximized with respect to $\mu$. 
 For comparison, those for the noncondensed state ($\theta$=0) are shown by the dashed lines.
The state P satisfying the condition $\omega=\mu$ corresponds to the point at $\rho_{\Sigma^-}/\rho_{\rm B}$=0.117 on the bold solid line. The state Q denoting the absolute energy minimum with $\theta$=0.70 rad corresponds to the point at $\rho_{\Sigma^-}/\rho_{\rm B}$=0 on the bold solid line. From comparison of each energy contribution at the states P and Q, one can see that the hyperon($Y$)-nucleon ($N$) mass difference mainly pushes up the total energy as the $\Sigma^-$-mixing ratio increases. 
The mixing of the $\Sigma^-$ hyperon slightly reduces the total kinetic energy of baryons by lowering the Fermi momentum of each baryon, while  slightly enlarging the baryon potential energy contribution.  
These two effects are compensated each other. The lepton energy contribution is litte changed by the $\Sigma^-$-mixing. 
Note that the sum of the kaon-baryon scalar interaction energy and free kaon energy consisting of the kaon free mass and kinetic energy is positive, and that it decreases as the $\Sigma^-$-mixing ratio increases. However, the decrease in the sum of the kaon-baryon scalar interaction energy and free kaon energy cannot compensate for the energy excess from the $Y$-$N$ mass difference as the $\Sigma^-$-mixing increases. It is to be noted that, for any value of the $\Sigma^-$-mixing ratio, all the energy contributions except for the sum of the kaon-baryon scalar interaction energy and free kaon energy have lower energy in the kaon-condensed state (solid lines) than in the noncondensed state (dashed lines).

The more detailed numerical analysis shows the following behavior for the energy minima in the ($\theta$, $\rho_{\Sigma^-}/\rho_B$) plane in the vicinity of $\rho_{\rm B}^{c(2)}(K^-)$ : At a certain density below $\rho_{\rm B}^{c(2)}(K^-)$, a local minimum state Q' corresponding to the $K^-$-condensed state without the $\Sigma^-$-mixing appears in addition to the absolute minimum state P' corresponding to the noncondensed state with the $\Sigma^-$-mixing. [We denote this density as $\rho_{\rm B}^\ast (K^-; {\rm no}\ \Sigma^-)$.] As the density increases, the state Q' shifts to have a lower energy, and at a density, denoted as $\rho_{\rm B}^{c(1)}(K^-; {\rm no} \ \Sigma^-)$, the energy values of the two minima P' and Q' get equal. Above $\rho_{\rm B}=\rho_{\rm B}^{c(1)}(K^-; {\rm no} \ \Sigma^-)$, the state Q' becomes the absolute minimum, having a lower energy than that of the state P'.  In Table~\ref{tab:onset}, we show the typical densities,  
$\rho_{\rm B}^\ast (K^-; {\rm no}\ \Sigma^-)$, 
$\rho_{\rm B}^{c(1)}(K^-; {\rm no} \ \Sigma^-)$ as well as $\rho_{\rm B}^{c(2)}(K^-)$. 
In the case of H-EOS (A) and $\Sigma_{Kn}$ = 305 MeV, there is a {\it discontinuous} transition from the noncondensed state of hyperonic matter with the $\Sigma^-$-mixing (the state P') to the $K^-$-condensed state without the $\Sigma^-$-mixing (the state Q') above $\rho_{\rm B}=\rho_{\rm B}^{c(1)}(K^-; {\rm no} \ \Sigma^-)$. This transition density $\rho_{\rm B}^{c(1)}(K^-; {\rm no} \ \Sigma^-)$ is slightly lower than $\rho_{\rm B}^{c(2)}(K^-)$.

\begin{table}[h]
\caption{The typical densities associated with the appearance of $K^-$ condensates and $\Sigma^-$ hyperons.  They are calculated with the EOS models for hyperonic matter, H-EOS (A) and H-EOS (B). 
The values in the parentheses for $\rho_{\rm B}^{c(2)}(K^-)$ mean that they don't correspond to the true energy minimum but the local minimum or the saddle point in the ($\theta$, $\rho_{\Sigma^-}/\rho_{\rm B}$) plane. See the text for details. }
\label{tab:onset}
\begin{center}
\begin{tabular}{c|c||c|c|c||c|c}
\hline
H-EOS & $\Sigma_{Kn}$ & $\rho_{\rm B}^\ast(K^-; {\rm no }\ \Sigma^-$)  & $\rho_{\rm B}^{c(1)}(K^-; {\rm no} \ \Sigma^-)$ & $\rho_{\rm B}^{c(2)}(K^-)$  & $\rho_{\rm B}^\ast(K^-; \Sigma^-)$  & $\rho_{\rm B}^{c(1)}(K^-; \Sigma^-)$ \\ 
 & (MeV) & (fm$^{-3}$) &   (fm$^{-3}$)  &  (fm$^{-3}$) & (fm$^{-3}$) &   (fm$^{-3}$)  \\\hline
 (A) & 305 & 0.5782 & 0.6135 & (0.6433) & 1.011 & 1.039   \\
 & 207 & 0.8280 & $-$ &  0.9254 & $-$ & $-$ \\\hline
(B) & 305 & $-$ & $-$ & 0.5504 & 1.006 & 1.069  \\
 & 207 & 0.7084 & 0.9086 &  (0.9189) & 0.9189 & 1.170 \\\hline
\end{tabular}
\end{center}
\end{table}

Next we proceed to the case of H-EOS (A) and $\Sigma_{Kn}$=207 MeV. As seen from Fig.~\ref{fig:contour-a}(b), the state P is an absolute minimum. Therefore, the assumption of the continuous transition is kept valid, and the onset density for kaon condensation is given by $ \rho_{\rm B}^{c(2)}(K^-)$. For this weaker kaon-baryon scalar attraction case, the critical density $\rho_{\rm B}^{c(2)}(K^-)$ is far beyond the onset density of $\Sigma^-$, $\rho_{\rm B}^c(\Sigma^-)$ (=0.52 fm$^{-3}$), so that the concentration of the $\Sigma^-$ hyperon in matter is not affected much by the appearance of kaon condensates. However, it should be noted that there still exists a kaon-condensed local minimum Q ($\theta$, $\rho_{\Sigma^-}/\rho_B$)=(1.0 rad, 0) without the $\Sigma^-$-mixing. Indeed, the local minimum (the state Q') exists from a fairly lower density $\rho_{\rm B}^\ast(K^-; {\rm no} \ \Sigma^-)$ (=0.8280 fm$^{-3}$) than $\rho_{\rm B}^{c(2)}(K^-)$ (=0.9254 fm$^{-3}$). [See Table~\ref{tab:onset}.]

\begin{figure}[h]
\begin{center}
\includegraphics[height=.3\textheight]{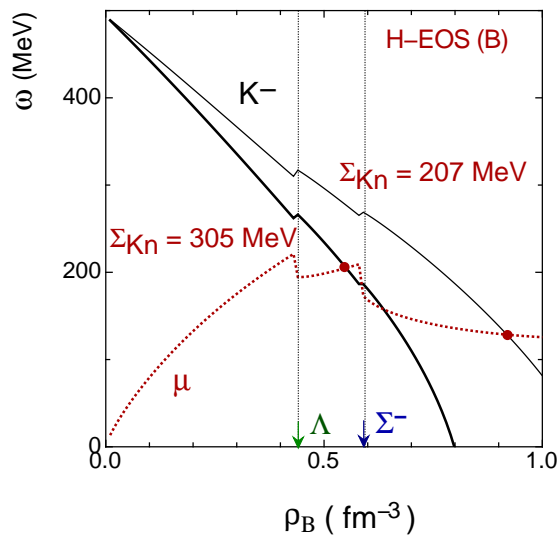}
\caption{The lowest energies of $K^-$ as functions of baryon number density $\rho_{\rm B}$ for $\Sigma_{Kn}$ = 305 MeV (bold solid line) and $\Sigma_{Kn}$ = 207 MeV (thin solid line) in the case of H-EOS (B). }
\label{fig:w-b}
\end{center}
\end{figure}

\begin{figure}[!]
\noindent\begin{minipage}[l]{0.50\textwidth}
\begin{center}
\includegraphics[height=.3\textheight]{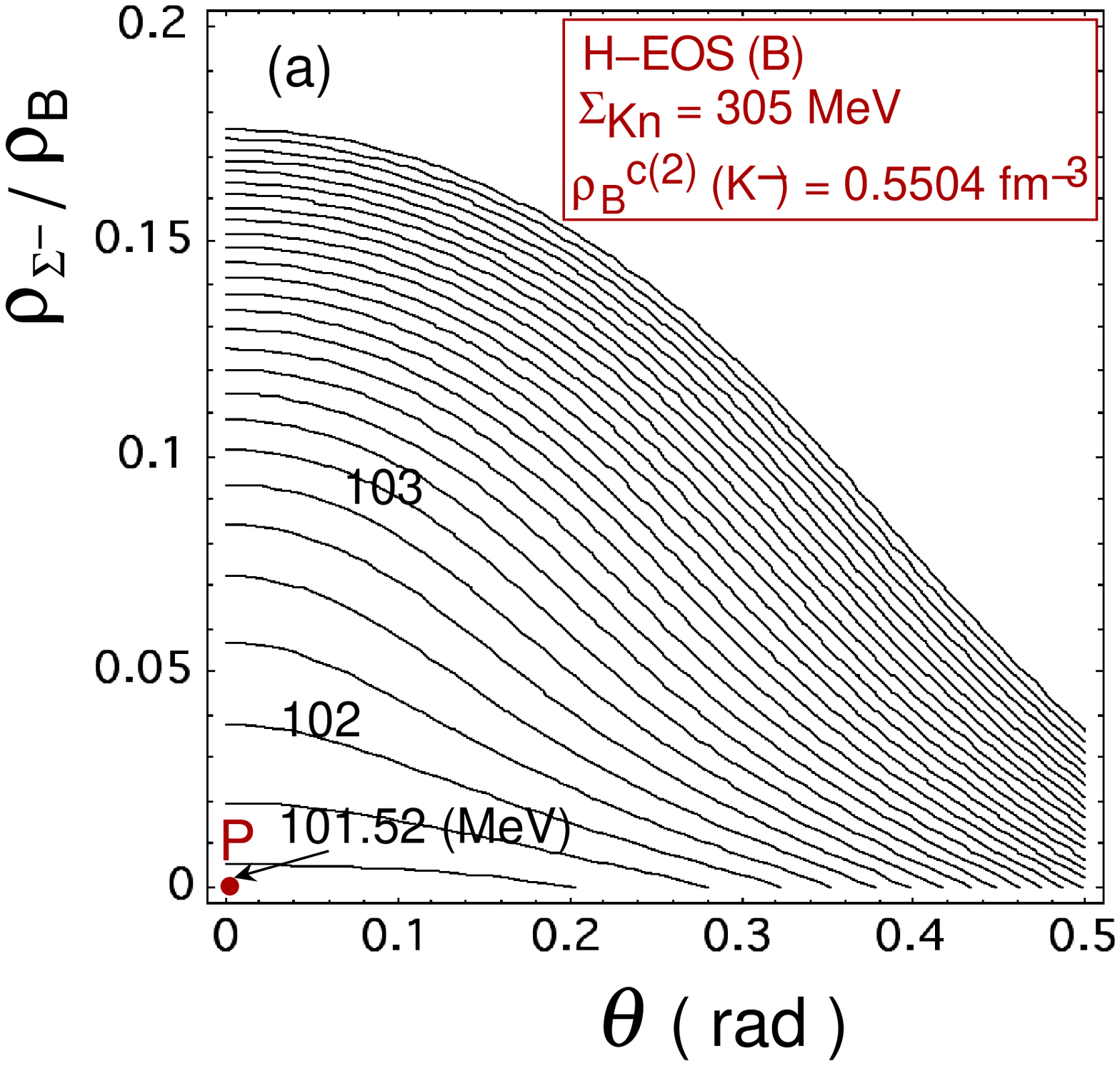}
\end{center}
\end{minipage}~
\begin{minipage}[r]{0.50\textwidth}
\begin{center}
\includegraphics[height=.3\textheight]{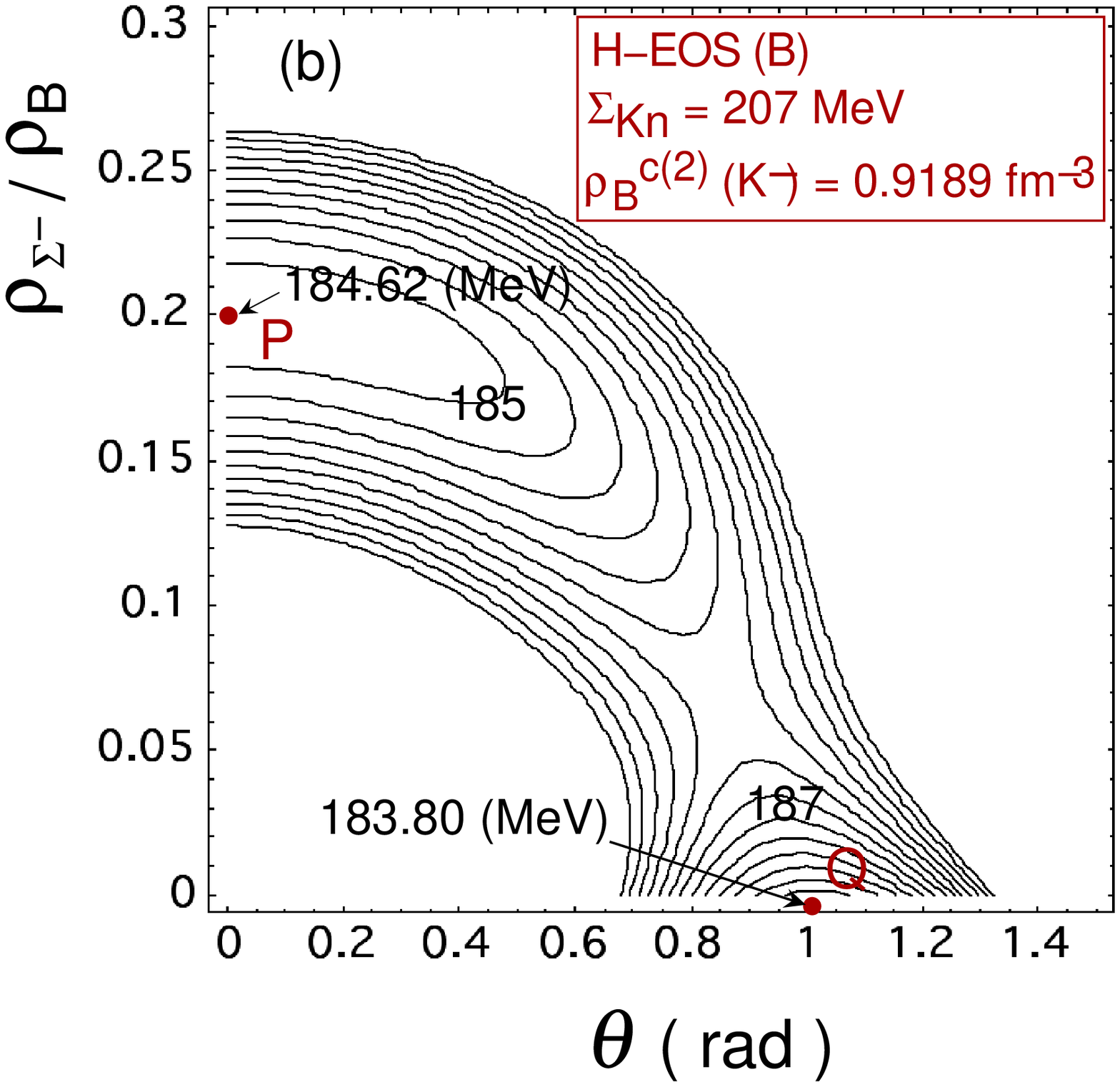}
\end{center}
\end{minipage}
\caption{(a) Contour plot of the total energy per baryon ${\cal E}'/\rho_{\rm B}$ in the ($\theta$, $\rho_{\Sigma^-}/\rho_B$) plane at $\rho_{\rm B}=\rho_{\rm B}^{c(2)}(K^-)$ for $\Sigma_{Kn}$=305 MeV in the case of H-EOS (B).  The energy interval is taken to be 0.2 MeV.  (b) The same as in (a), but for $\Sigma_{Kn}$=207 MeV. The energy interval is taken to be 0.5 MeV. See the text for details. }
\label{fig:contour-b}
\end{figure}

\subsection{Case of H-EOS (B) }
\label{subsec:b}

\ \ In the case of H-EOS (B) for the hyperonic matter EOS, both the $\Lambda$ and $\Sigma^-$ start to be mixed at higher densities as compared with the case of H-EOS (A)
 [Sec.~\ref{sec:fraction}].  
In Fig.~\ref{fig:w-b}, we show the lowest energies of the $K^-$ as functions of baryon number density $\rho_{\rm B}$ for $\Sigma_{Kn}$ = 305 MeV (bold solid line) and $\Sigma_{Kn}$ = 207 MeV (thin solid line). 
In Fig.~\ref{fig:contour-b}, the 
contour plots of the total energy per baryon ${\cal E}'/\rho_{\rm B}$ 
in the ($\theta$, $\rho_{\Sigma^-}/\rho_B$) plane at $\rho_{\rm B}=\rho_{\rm B}^{c(2)}(K^-)$ are depicted for  
$\Sigma_{Kn}$=305 MeV [Fig.~\ref{fig:contour-b}(a)] and $\Sigma_{Kn}$=207 MeV [Fig.~\ref{fig:contour-b}(b)] in the case of H-EOS (B). 
From Fig.~\ref{fig:w-b}, the critical density $\rho_{\rm B}^{c(2)}(K^-)$ is read as $\rho_{\rm B}^{c(2)}(K^-)$=0.5504~fm$^{-3}$ (=3.44$\rho_0$) for $\Sigma_{Kn}$=305 MeV and $\rho_{\rm B}^{c(2)}(K^-)$=0.9189~fm$^{-3}$ (=5.74$\rho_0$) for $\Sigma_{Kn}$=207 MeV. 

For $\Sigma_{Kn}$=305 MeV, the condition $\omega=\mu$ [Eq.~(\ref{eq:onset}) ] is satisfied before mixing of the $\Sigma^-$ starts, i.e., $\rho_{\rm B}^{c(2)}(K^-) < \rho_{\rm B}^{c}(\Sigma^-)$. As seen from Fig.~\ref{fig:contour-b}(a), the corresponding state P is the  absolute minimum for 
the total energy per baryon ${\cal E}'/\rho_{\rm B}$ in the ($\theta$, $\rho_{\Sigma^-}/\rho_B$) plane at $\rho_{\rm B}=\rho_{\rm B}^{c(2)}(K^-)$, and there is no local minimum with kaon condensates without the $\Sigma^-$-mixing. 
Therefore, the assumption of the continuous phase transition does not lose its validity when the $\Sigma^-$ hyperons are not mixed in the ground state. 

On the other hand, for $\Sigma_{Kn}$=207 MeV, the state P  satisfying the condition $\omega=\mu$ is obtained as a local minimum at a point ($\theta$, $\rho_{\Sigma^-}/\rho_{\rm B}$)=(0, 0.200) in the presence of the $\Sigma^-$, and there is an absolute minimum with kaon condensates without the $\Sigma^-$-mixing (the state Q in Fig.~\ref{fig:contour-b}(b)) at a point ($\theta$, $\rho_{\Sigma^-}/\rho_{\rm B}$)=(1.02 rad, 0). As compared with the case of H-EOS (A), the critical density $\rho_{\rm B}^{c(2)}(K^-)$ is not so far from the onset density of the $\Sigma^-$, $\rho_{\rm B}^c(\Sigma^-)$. [ $\rho_{\rm B}^{c(2)}(K^-)-\rho_{\rm B}^c(\Sigma^-)$ = 2.1 $\rho_0$ for H-EOS (B), while $\rho_{\rm B}^{c(2)}(K^-)-\rho_{\rm B}^c(\Sigma^-)$ = 2.5 $\rho_0$ for H-EOS (A). ]
As a result, competition between the $\Sigma^-$ and $K^-$ condensates is more remarkable in the case of H-EOS (B) and $\Sigma_{Kn}$ = 207 MeV than in the case of H-EOS (A) and $\Sigma_{Kn}$ = 207 MeV, making the state Q energetically more favorable than the state P. From Table~\ref{tab:onset}, one can see the common behavior as the case of H-EOS (A) and $\Sigma_{Kn}$ = 305 MeV concerning the appearance of $K^-$ condensates and $\Sigma^-$ hyperons : 
There is a {\it discontinuous} transition from the noncondensed state of hyperonic matter with the $\Sigma^-$-mixing to the $K^-$-condensed state without the $\Sigma^-$-mixing above $\rho_{\rm B}=\rho_{\rm B}^{c(1)}(K^-; {\rm no} \ \Sigma^-)$, and this transition density $\rho_{\rm B}^{c(1)}(K^-; {\rm no} \ \Sigma^-)$ is slightly lower than $\rho_{\rm B}^{c(2)}(K^-)$.

\section{Equation of State}
\label{sec:eos}

\begin{figure}[h]
\noindent\begin{minipage}[l]{0.50\textwidth}
\begin{center}
\includegraphics[height=.3\textheight]{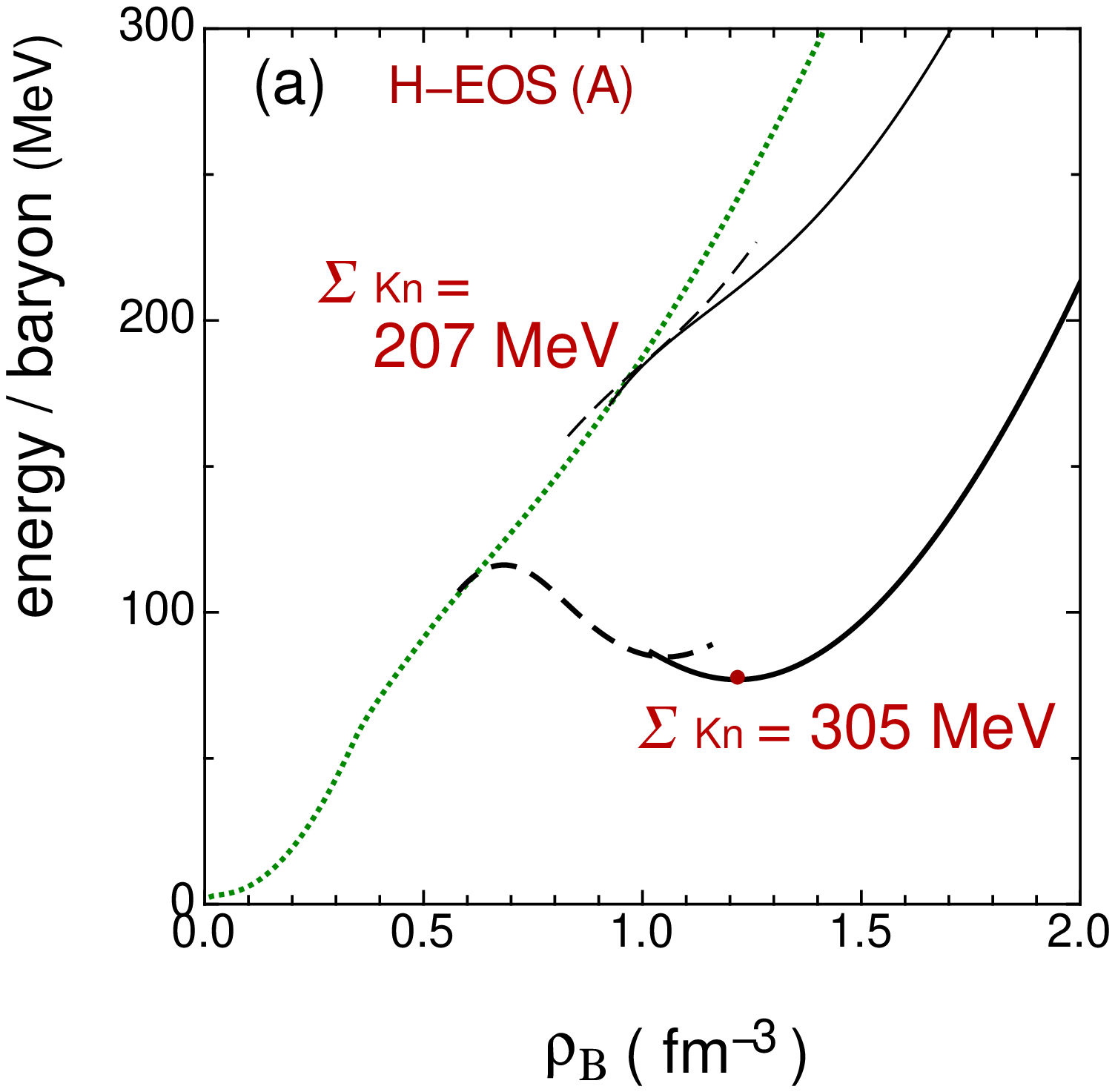}
\end{center}
\end{minipage}~
\begin{minipage}[r]{0.50\textwidth}
\begin{center}
\includegraphics[height=.3\textheight]{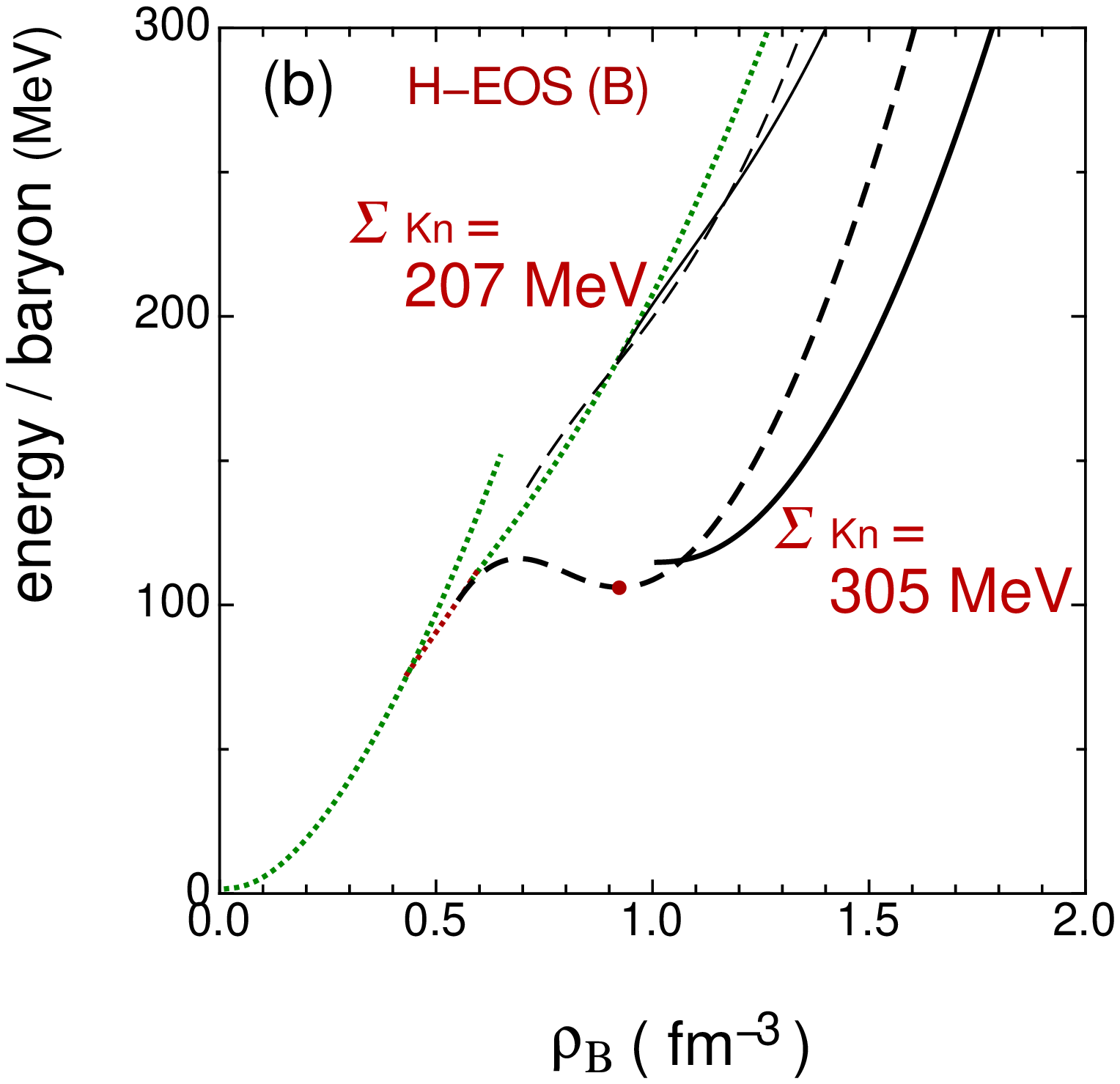}
\end{center}
\end{minipage}
\caption{(a) The total energies per baryon in the $K^-$-condensed phase, ${\cal E}'/\rho_{\rm B}$, as functions of the baryon number density $\rho_{\rm B}$ for H-EOS (A). The bold (thin) lines are for $\Sigma_{Kn}$ = 305 MeV ($\Sigma_{Kn}$ = 207 MeV). The solid lines stand for the total energies per baryon for the $K^-$-condensed state with the $\Sigma^-$-mixing, while the dashed lines for the $K^-$- condensed state without the $\Sigma^-$-mixing. For comparison, the energy per baryon for the noncondensed hyperonic matter [H-EOS (A)] is shown by the dotted line.
(b) The same as in (a), but for H-EOS (B).}
\label{fig:eos}
\end{figure}

\begin{figure}[!]
\noindent\begin{minipage}[l]{0.50\textwidth}
\begin{center}
\includegraphics[height=.3\textheight]{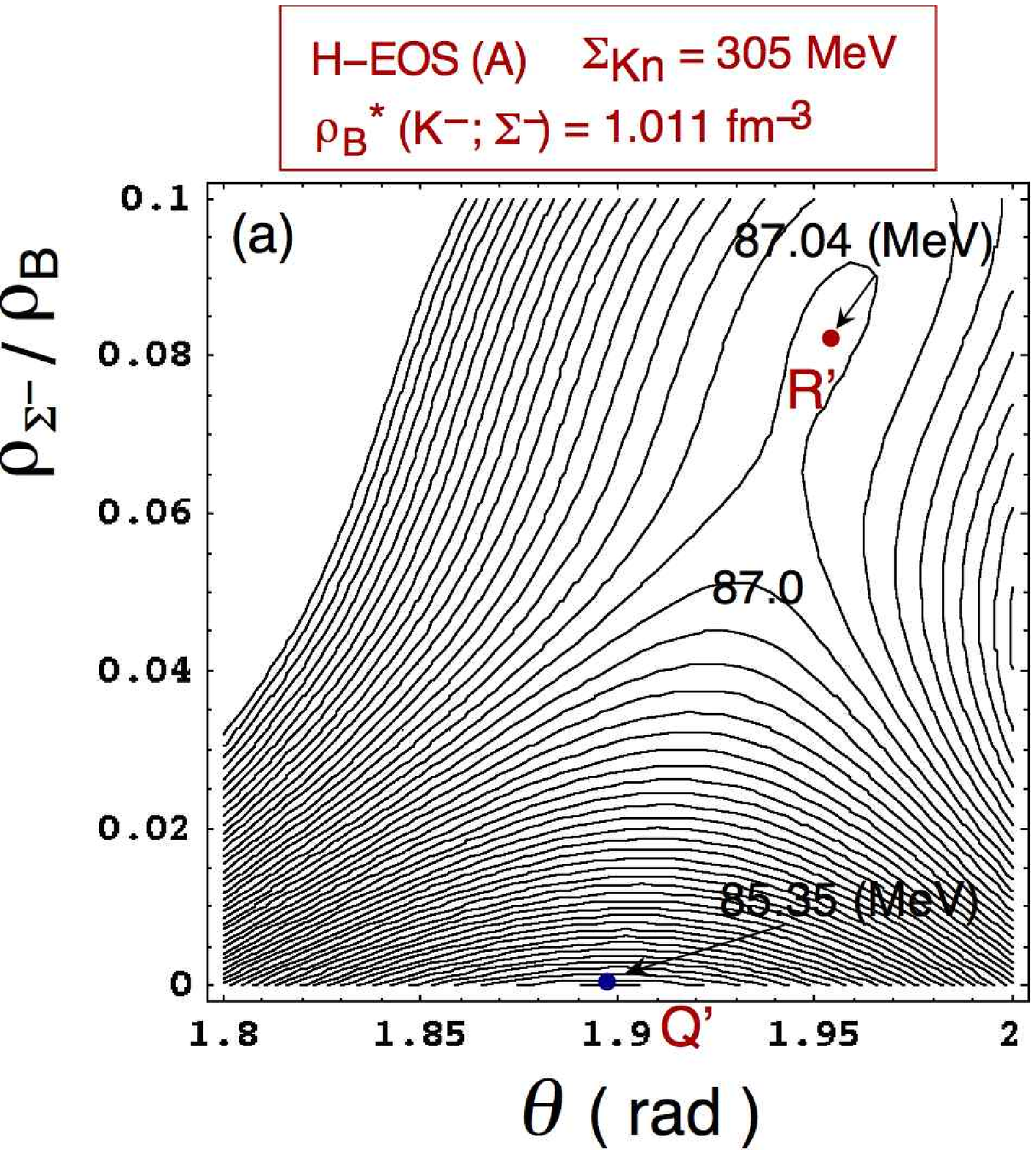}
\end{center}
\end{minipage}~
\begin{minipage}[r]{0.50\textwidth}
\begin{center}
\includegraphics[height=.3\textheight]{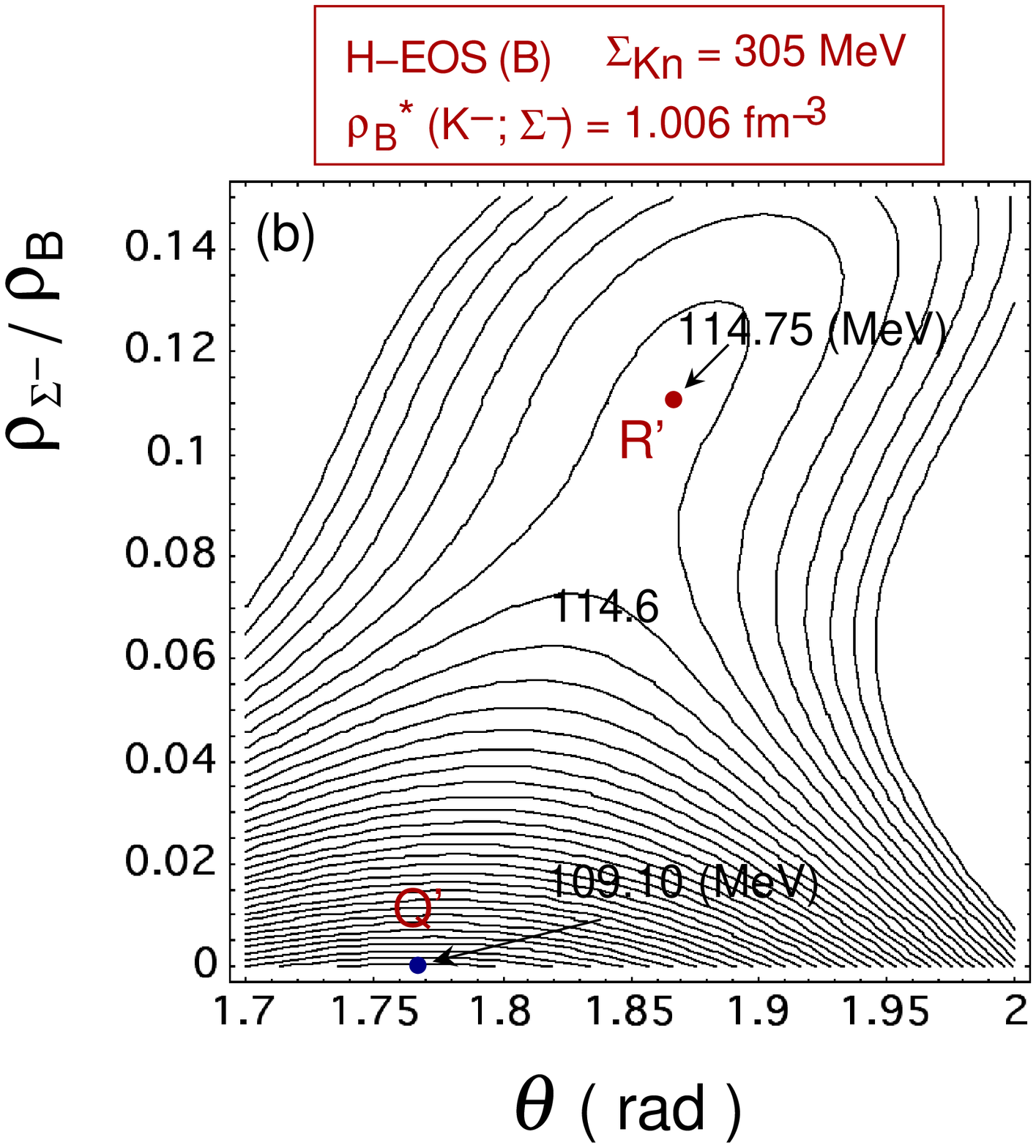}
\end{center}
\end{minipage}
\caption{(a) Contour plot of the total energy per baryon ${\cal E}'/\rho_{\rm B}$ in the ($\theta$, $\rho_{\Sigma^-}/\rho_B$) plane at $\rho_{\rm B}=\rho_{\rm B}^\ast(K^-; \Sigma^-)$ for $\Sigma_{Kn}$=305 MeV in the case of H-EOS (A).  The energy interval is taken to be 0.05 MeV.  (b) The same as in (a), but in the case of H-EOS (B). The energy interval is taken to be 0.2 MeV. See the text for details.}
\label{fig:contour2}
\end{figure}

\begin{figure}[!]
\noindent\begin{minipage}[l]{0.50\textwidth}
\begin{center}
\includegraphics[height=.3\textheight]{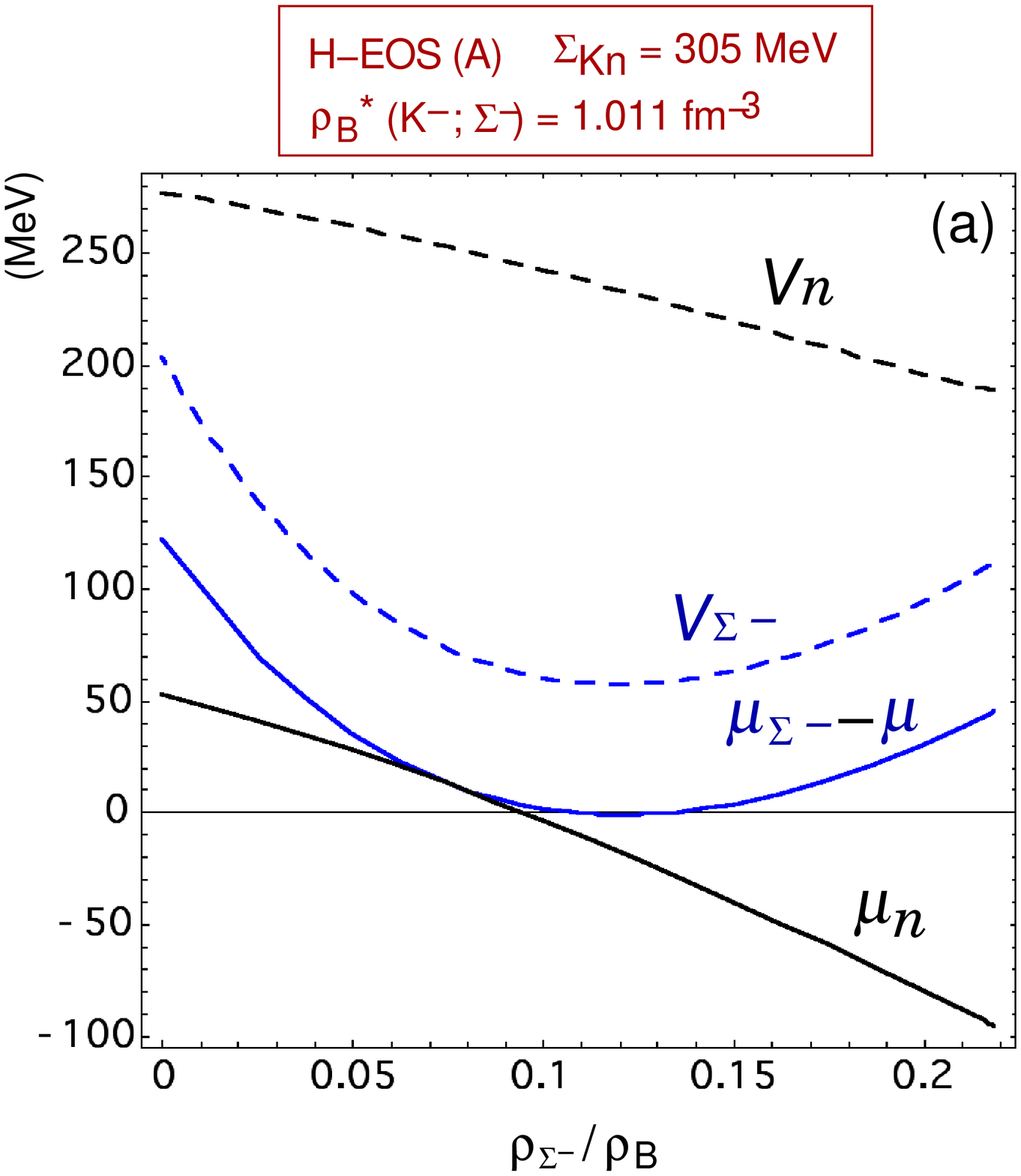}
\end{center}
\end{minipage}~
\begin{minipage}[r]{0.50\textwidth}
\begin{center}
\includegraphics[height=.3\textheight]{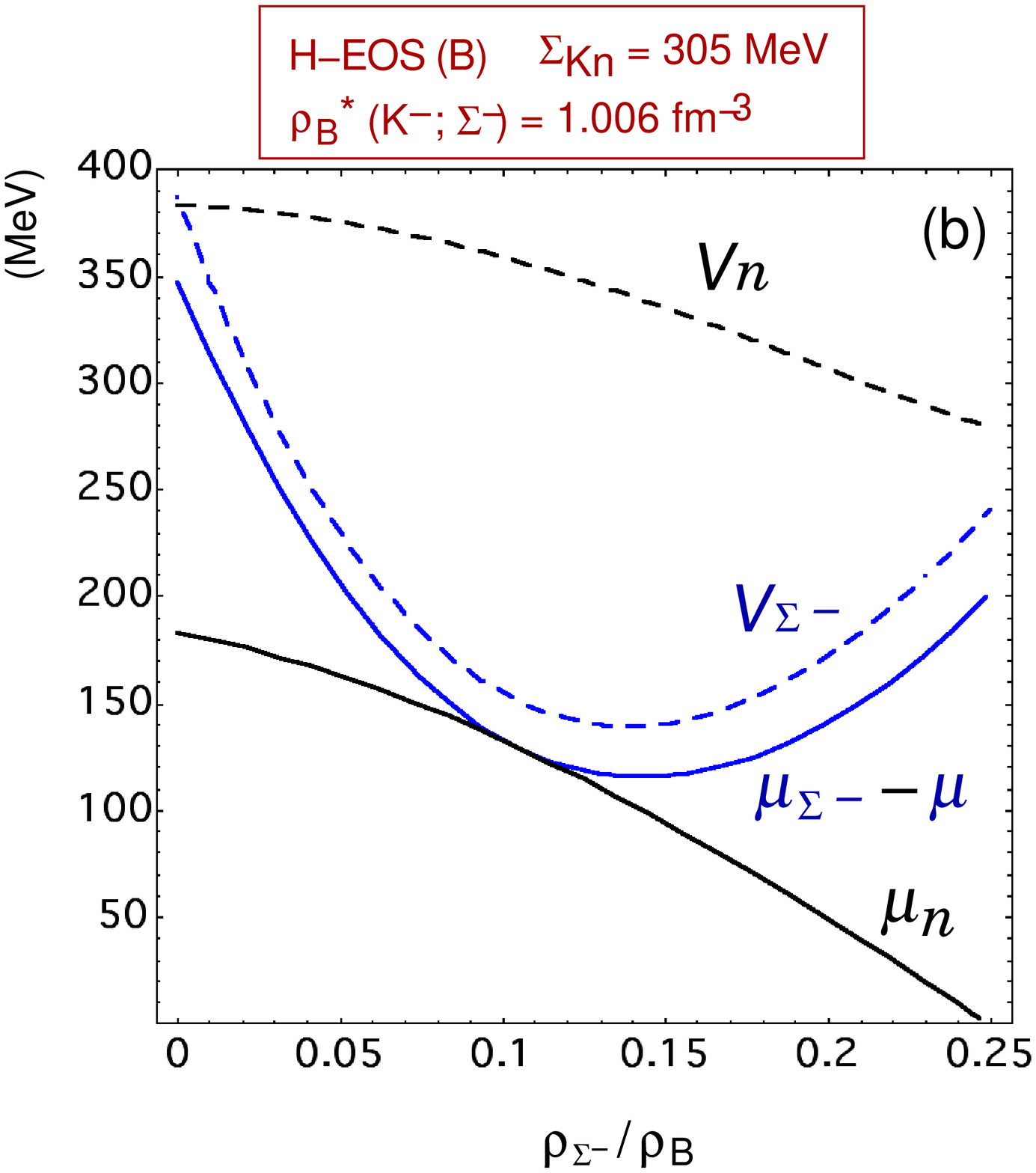}
\end{center}
\end{minipage}
\caption{(a) Dependence of $V_{\Sigma^-}$, $V_n$, and $\mu_{\Sigma^-}-\mu$, $\mu_n$ on the $\Sigma^-$-mixing ratio $\rho_{\Sigma^-}/\rho_{\rm B}$ in the kaon-condensed phase at $\rho_{\rm B}=\rho_{\rm B}^\ast(K^-; \Sigma^-)$ for $\Sigma_{Kn}$=305 MeV with H-EOS (A). (b) The same as in (a), but in the case of H-EOS (B). }
\label{fig:cep}
\end{figure}

\subsection{Two energy minima with and without the $\Sigma^-$-mixing for the $K^-$-condensed phase}
\label{subsubsec:two-minima}

\ \ Here we discuss the EOS of the $K^-$-condensed phase in hyperonic matter. The total energies per baryon in the $K^-$-condensed phase, ${\cal E}'/\rho_{\rm B}$, as functions of the baryon number density $\rho_{\rm B}$ are shown in Fig.~\ref{fig:eos}. 
Fig.~\ref{fig:eos} (a) is for H-EOS (A), and (b) is for H-EOS (B). 
The bold (thin) lines are for $\Sigma_{Kn}$ = 305 MeV ($\Sigma_{Kn}$ = 207 MeV). The solid lines stand for the total energies per baryon for the $K^-$-condensed state with the $\Sigma^-$-mixing, while the dashed lines for the $K^-$-condensed state without the $\Sigma^-$-mixing. For comparison, the energy per baryon for the noncondensed hyperonic matter is shown by the dotted line.
In each case of the model EOS for hyperonic matter and the $Kn$ sigma term $\Sigma_{Kn}$, there are two solutions of the kaon-condensed phase corresponding to two minima in the ($\theta$, $\rho_{\Sigma^-}/\rho_{\rm B}$) plane at some density intervals: One is the $K^-$-condensed state without the $\Sigma^-$-mixing (dashed lines) called the state Q' in Sec.~\ref{sec:validity}, and the other is with the $\Sigma^-$-mixing (solid lines), which we call the state R'. The density at which the state R' appears as a local minimum is denoted as $\rho_{\rm B}^\ast(K^-; \Sigma^-)$. 
For example, we show the contour plots of the total energy per baryon ${\cal E}'/\rho_{\rm B}$ 
in the ($\theta$, $\rho_{\Sigma^-}/\rho_{\rm B}$) plane at $\rho_{\rm B}=\rho_{\rm B}^\ast(K^-; \Sigma^-)$ for $\Sigma_{Kn}$=305 MeV in the case of H-EOS (A) in Fig.~\ref{fig:contour2}(a) and H-EOS (B) in Fig.~\ref{fig:contour2}(b). 
In Fig.~\ref{fig:cep}, we also show the dependence of the baryon potentials $V_{\Sigma^-}$, $V_n$, the neutron chemical potential $\mu_n$, and the difference of the $\Sigma^-$ and charge chemical potentials, $\mu_{\Sigma^-}-\mu$, on the $\Sigma^-$-mixing ratio $\rho_{\Sigma^-}/\rho_{\rm B}$ in the kaon-condensed phase at $\rho_{\rm B}=\rho_{\rm B}^\ast(K^-; \Sigma^-)$. Fig.~\ref{fig:cep}(a) is for $\Sigma_{Kn}$=305 MeV with H-EOS (A) and Fig.~\ref{fig:cep}(b) is for $\Sigma_{Kn}$=305 MeV  with H-EOS (B).
One can see that the chemical potential difference $\mu_{\Sigma^-}-\mu$ has a minimum at a finite value of $\rho_{\Sigma^-}/\rho_{\rm B}$ (= 0.12 $-$ 0.14) and that the neutron chemical potential $\mu_n$ decreases monotonically with increase in the $\Sigma^-$-mixing ratio within the range $ \rho_{\Sigma^-}/\rho_{\rm B}$ = 0.0 $-$ 0.3. 
As a result, the chemical equilibrium condition, $\mu_n=\mu_{\Sigma^-}-\mu$, for the weak process $ne^-\rightleftharpoons \Sigma^-(\nu_e)$ is met at a finite $\Sigma^-$-mixing ratio (= 0.07 $-$ 0.1), which corresponds to the appearance of the state R' in Fig.~\ref{fig:contour2}. The dependence of the chemical potentials $\mu_{\Sigma^-}-\mu$ and $\mu_n$ on the $\Sigma^-$-mixing ratio is caused  from that of the baryon potentials $V_{\Sigma^-}$ and $V_n$, respectively, as seen from Fig.~\ref{fig:cep}.
 
 As the density increases, the difference of the energies between the state R' and the absolute minimum state Q' gets smaller, and at a certain density denoted as $\rho_{\rm B}^{c(1)}(K^-; \Sigma^-)$, the energies of the states Q' and R' get equal. Above the density $\rho_{\rm B}^{c(1)}(K^-; \Sigma^-)$, the state R' develops as an absolute energy minimum. In Table~\ref{tab:onset}, we show the numerical values of $ \rho_{\rm B}^\ast(K^-; \Sigma^-)$ and $\rho_{\rm B}^{c(1)}(K^-; \Sigma^-)$ for each case of $\Sigma_{Kn}$ and the hyperonic matter EOS. 
 
At a given density, the ground state is determined by the lowest energy state in Fig.~\ref{fig:eos}. For all the cases, the state R' becomes the ground state at higher densities, i.e., the $\Sigma^-$ is mixed in the fully-developed kaon-condensed phase. Except for the case of $\Sigma_{Kn}$=207 MeV with H-EOS (A), the transition from the Q' state (the dashed lines) to the R' state (the solid lines) is discontinuous. 
Even when the state R' is the ground state, the local minimum (the state Q') prevails over the wide range of the baryon number density, in particular, in the case of H-EOS (B) [see the dashed lines in Fig.~\ref{fig:eos} (b)]. 

\begin{figure}[h]
\noindent\begin{minipage}[l]{0.50\textwidth}
\begin{center}
\includegraphics[height=.3\textheight]{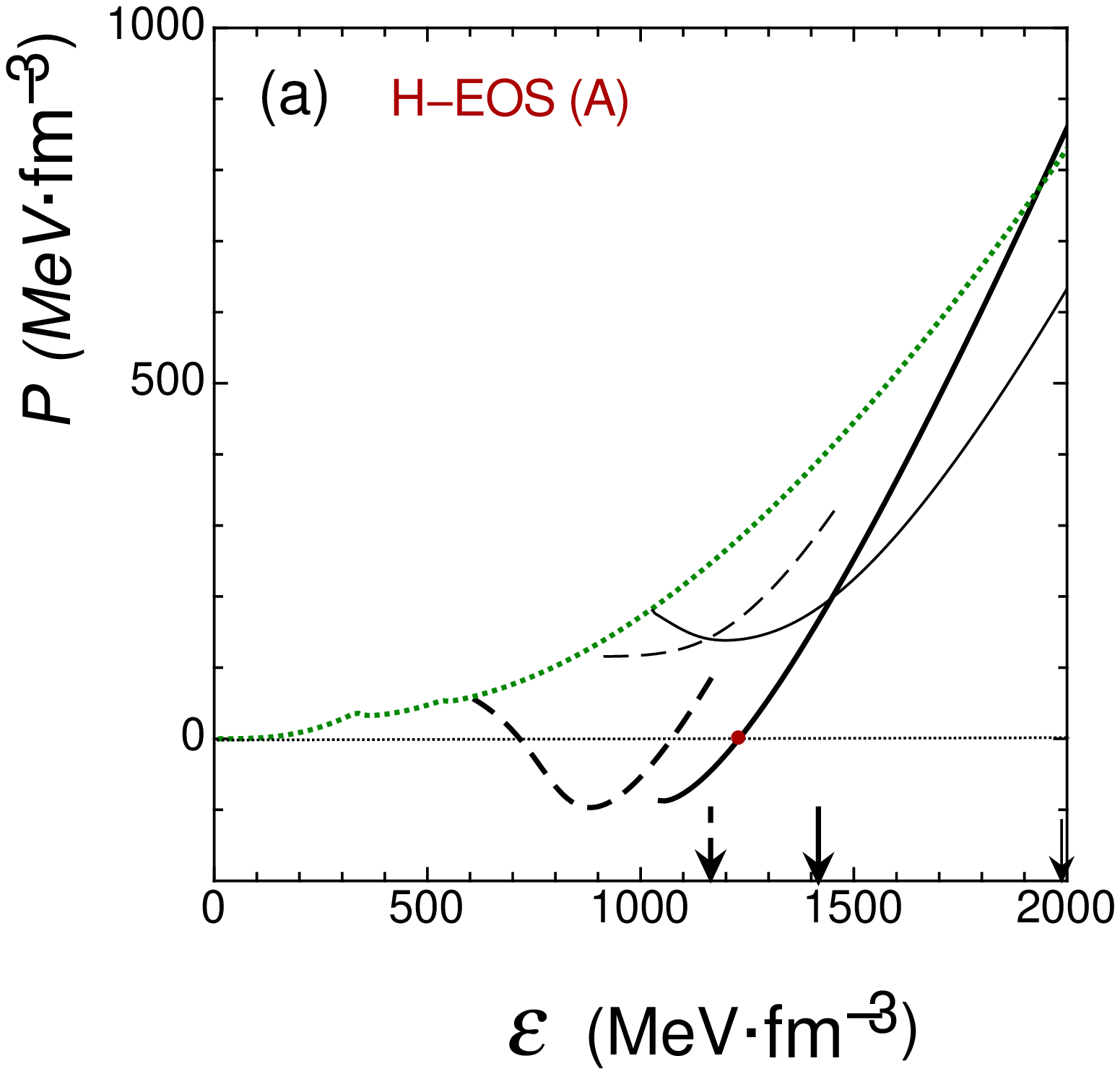}
\end{center}
\end{minipage}~
\begin{minipage}[r]{0.50\textwidth}
\begin{center}
\includegraphics[height=.3\textheight]{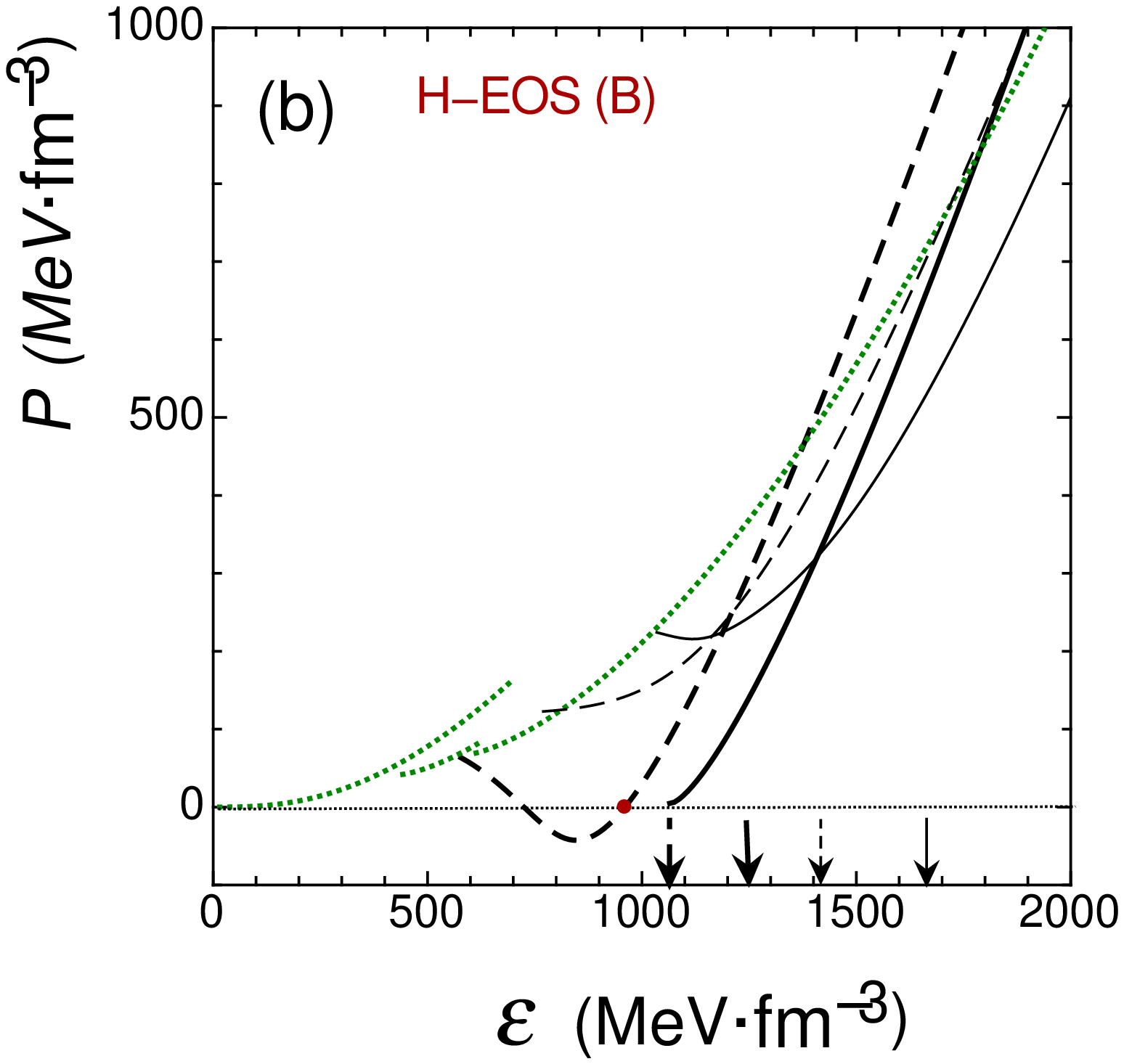}
\end{center}
\end{minipage}
\caption{(a) Pressure (MeV$\cdot$fm$^{-3}$) in the $K^-$-condensed phase, $P\equiv \rho_{\rm B}^2\partial({\cal E}'/\rho_{\rm B})/\partial \rho_{\rm B}$ (=$-{\cal E}_{\rm eff}'$), as functions of the energy density $\epsilon$ (MeV$\cdot$fm$^{-3}$) for H-EOS (A). The notations of the lines are the same as those in Fig.~8.  
(b) The same as in (a), but for H-EOS (B). }
\label{fig:pres}
\end{figure}

In Fig.~\ref{fig:pres}, we show the pressure in the $K^-$-condensed phase obtained by $P\equiv \rho_{\rm B}^2\partial({\cal E}'/\rho_{\rm B})/\partial \rho_{\rm B}$ (=$-{\cal E}_{\rm eff}'$), as functions of the energy density, $\epsilon\equiv{\cal E}'+M_N\rho_{\rm B}$, in MeV$\cdot$fm$^{-3}$. 
Fig.~\ref{fig:pres} (a) is for H-EOS (A), and (b) is for H-EOS (B). 
The bold (thin) lines are for $\Sigma_{Kn}$ = 305 MeV ($\Sigma_{Kn}$ = 207 MeV). The solid lines stand for the pressure for the $K^-$-condensed state with the $\Sigma^-$-mixing, while the dashed lines for the $K^-$-condensed state without the $\Sigma^-$-mixing. For comparison, the pressure for the noncondensed hyperonic matter is shown by the dotted line. Except for the case of $\Sigma_{Kn}$ = 207 MeV with H-EOS (A), there appears a gap in the pressure at the transition density $\rho_{\rm B}^{c(1)}(K^-; \Sigma^-)$ as a result of the discontinuous transition. It should be noted that the sound speed, 
$(\partial P/\partial\epsilon)^{1/2}$, exceeds the speed of light $c$ above a certain energy density, which is denoted as an arrow for each corresponding pressure curve. In such high density region, the relativistically covariant formulation is necessary for quantitative discussion of the EOS.  

\subsection{Density isomer state in the case of the stronger $s$-wave kaon-baryon scalar attraction}
\label{subsec:dis}

\ \ In the kaon-condensed phase realized in hyperonic matter, the EOS becomes considerably soft. In particular, in the case of $\Sigma_{Kn}$ = 305 MeV for both H-EOS (A) and (B), there appears a local energy minimum (which we call the density isomer state) at a certain density $\rho_{\rm B, min}$, and the pressure becomes negative at some density intervals below $\rho_{\rm B, min}$, as seen in Fig.~\ref{fig:eos} and \ref{fig:pres} (bold lines). For H-EOS (A) [H-EOS (B)], one reads $\rho_{\rm B, min}$ = 1.22 fm$^{-3}$ (0.92 fm$^{-3}$), and the minimum energy per baryon at $\rho_{\rm B, min}$ is 76.9 MeV (106.1 MeV), which is smaller than the $\Lambda$-$N$ mass difference, $\delta M_{\Lambda N}$=176 MeV. Thus the density isomer state is stable against the strong decay processes.  
 
\begin{figure}[h]
\noindent\begin{minipage}[l]{0.50\textwidth}
\begin{center}
\includegraphics[height=.3\textheight]{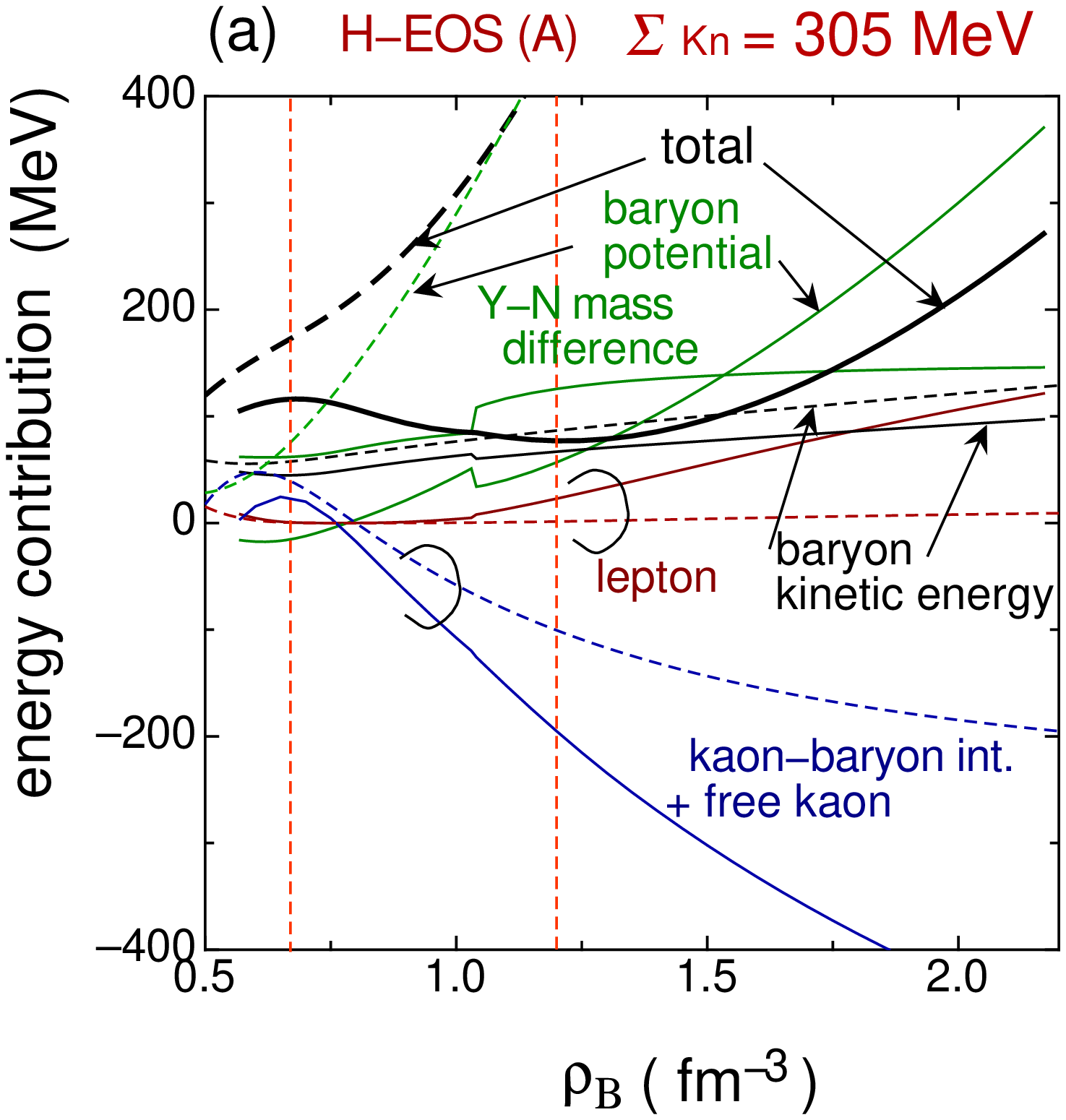}
\end{center}
\end{minipage}~
\begin{minipage}[r]{0.50\textwidth}
\begin{center}
\includegraphics[height=.3\textheight]{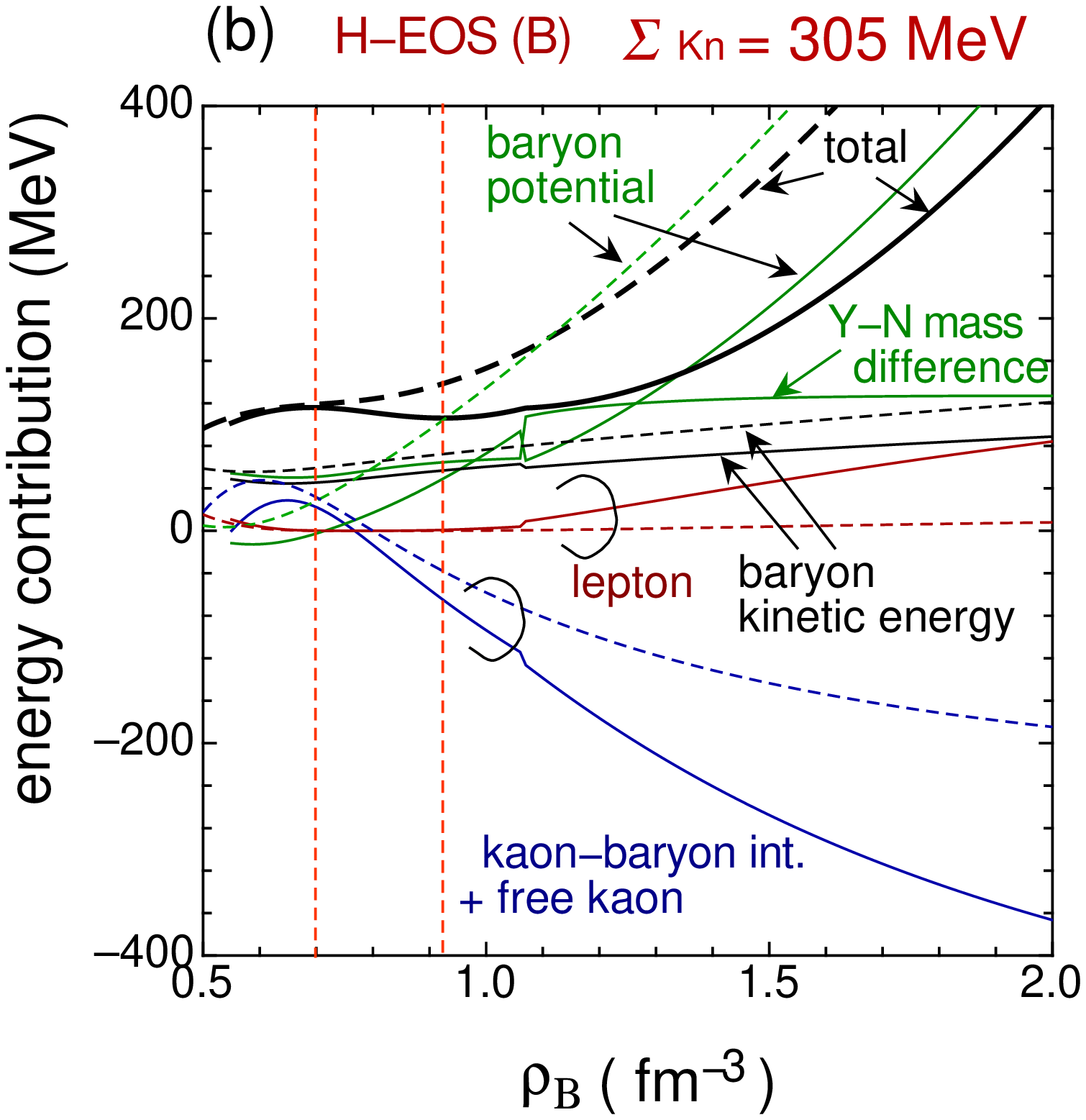}
\end{center}
\end{minipage}
\caption{(a) The contributions to the total energy per baryon for the kaon-condensed phase in hyperonic matter as functions of the baryon number density $\rho_{\rm B}$ for H-EOS (A) and $\Sigma_{Kn}$ = 305 MeV (solid lines). For comparison, those for the kaon-condensed phase realized in ordinary neutron-star matter, obtained after putting $\rho_\Lambda$=$\rho_{\Sigma^-}$=0, are shown by the dashed lines. 
(b) The same as in (a), but for H-EOS (B). See the text for details.}
\label{fig:e-contrib}
\end{figure}

 In order to clarify mechanisms for the significant softening of the EOS leading to the appearance of the local energy minimum and for subsequent recovering of the stiffness of the EOS at higher density region, we show the energy contributions to the total energy per baryon by the solid lines in the case of $\Sigma_{Kn}$ = 305 MeV for H-EOS (A) and H-EOS (B) in Figs.~\ref{fig:e-contrib}~(a) and (b), respectively. 
 For comparison, those for the kaon-condensed phase realized in ordinary neutron-star matter, obtained after putting $\rho_\Lambda$=$\rho_{\Sigma^-}$=0, are shown by the dashed lines. 
 The density region where the total energy per baryon decreases with density (i.e., the negative pressure region) is bounded by the vertical dotted lines in Figs.~\ref{fig:e-contrib} (a) and (b). 
The dependence of the total energy on the baryon number density is mainly determined by the two contributions: (I) the contribution from the classical kaons as the sum of the $s$-wave scalar kaon-baryon interaction and the free parts of the condensed kaon energy [the fourth, fifth and sixth terms in Eq.~(\ref{eq:te2})] and (II) the baryon potential energy ${\cal E}_{\rm pot}/\rho_{\rm B}$ 
[the third term in Eq.~(\ref{eq:te2})]. 
The contribution (I) decreases with increase in density, while the contribution (II) increases as density increases. 
As one can see from comparison of the solid lines 
and the dashed lines in Fig.~\ref{fig:e-contrib}, 
 the attractive effect from the contribution (I) is pronounced due to mixing of hyperons as compared with the case without hyperons, lowering the total energy at a given density. In addition, the repulsive effect from the contribution (II) is much weakened due to mixing of hyperons at a given density, since the repulsive interaction between nucleons is avoided by lowering the relative nucleon density through mixing of hyperons.\footnote{This suppression mechanism of the repulsive interaction between nucleons is essentially the same as that for softening of the EOS in the noncondensed hyperonic matter as pointed out in Ref.~\cite{y02,t04}.} 
As a result, the total energy is much reduced, leading to  significant softening of the EOS for the kaon-condensed phase in hyperonic matter. 
 
For the density region bounded by the dotted lines in Figs.~\ref{fig:e-contrib} (a) and (b), the increase of the absolute value of the kaon-baryon attractive interaction with density is more remarkable than the increase of the potential energy with density, so that the total energy per baryon decreases with density in this density region, and there is an energy minimum at $\rho_{\rm B, min}$. At higher densities above $\rho_{\rm B, min}$, the decrease of the contribution (I) gets slightly moderate, while the repulsive interaction between baryons becomes strong and so the increase of the contribution (II) with density gets more marked. As a result, the total energy per baryon increases rapidly with density, and the EOS recovers the stiffness at high densities. Thus the stiffness of the EOS depends on the quantitative behavior of the repulsive interaction between baryons at high densities, which has ambiguity depending on the model interaction. In our framework, the H-EOS (B) brings about stiffer EOS at high densities than the H-EOS (A), since the many-body hyperon-nucleon and hyperon-hyperon repulsive interaction terms, which control the stiffness of the EOS at high densities, contribute more significantly for H-EOS (B) [the index $\gamma$=2.0] than for H-EOS(A) [$\gamma$=5/3]. 

It has been suggested in Ref.~\cite{m05} that a density isomer state with kaon-condensates in hyperonic matter implies the existence of self-bound objects, which can be bound essentially without gravitation on a scale of an atomic nucleus to a neutron star just like a strangelet and a strange star\cite{b71,ck79,w84,fj84} or other exotic  matter\cite{lw74,h75,m76,lnt90,shsg02}. The density isomer state is located at a local energy minimum with respect to  the baryon number density as a metastable state, but it decays only through multiple weak processes, so that it is regarded as substantially stable.  
Implications of such self-bound objects with kaon condensates for astrophysical phenomena and nuclear experiments will be discussed in detail in a subsequent paper, where both $s$-wave and $p$-wave kaon-baryon interactions are taken into account\cite{m06}.

\begin{figure}[!]
\noindent\begin{minipage}[l]{0.50\textwidth}
\begin{center}
\includegraphics[height=.3\textheight]{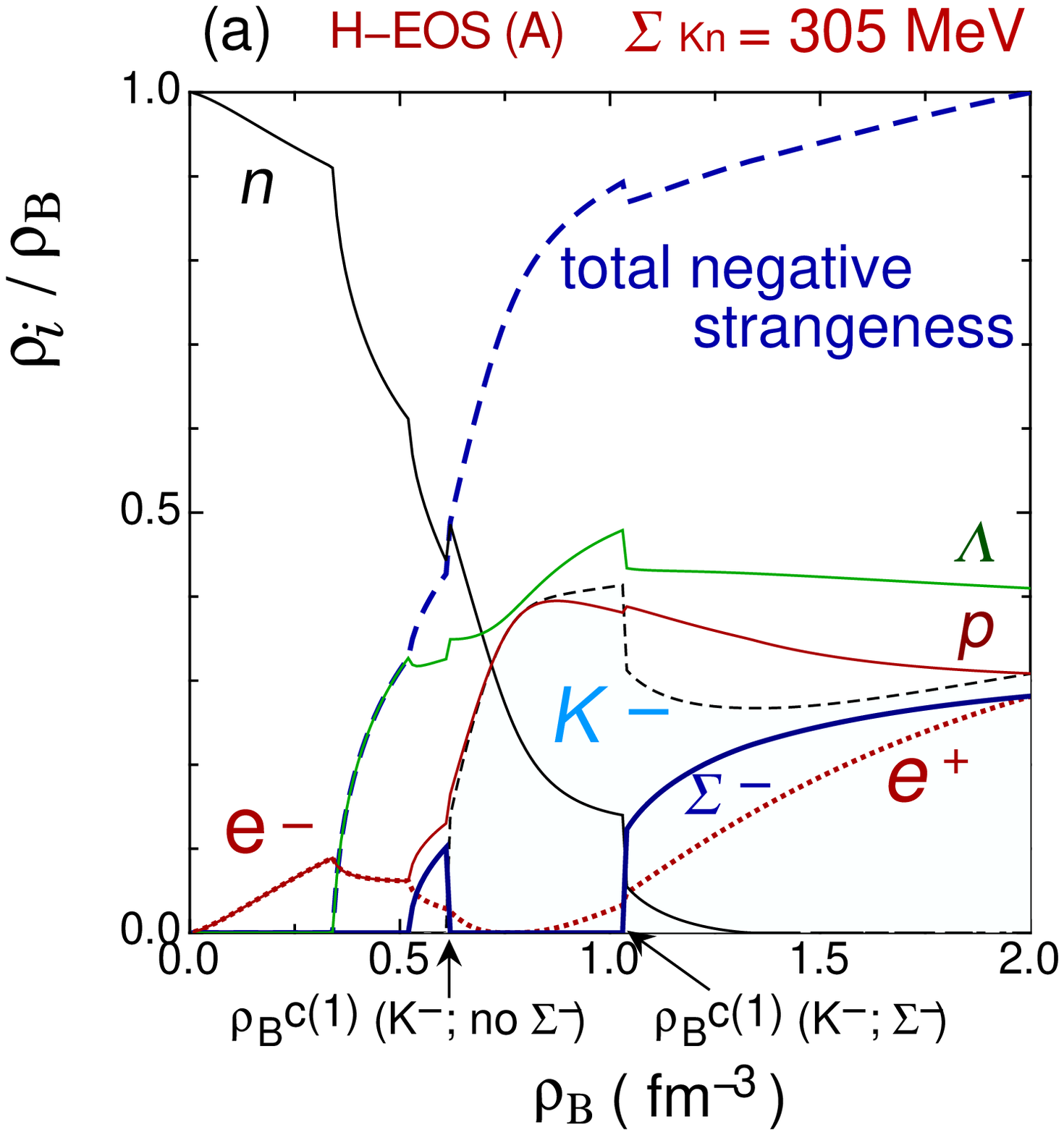}
\end{center}
\end{minipage}~
\begin{minipage}[r]{0.50\textwidth}
\begin{center}
\includegraphics[height=.3\textheight]{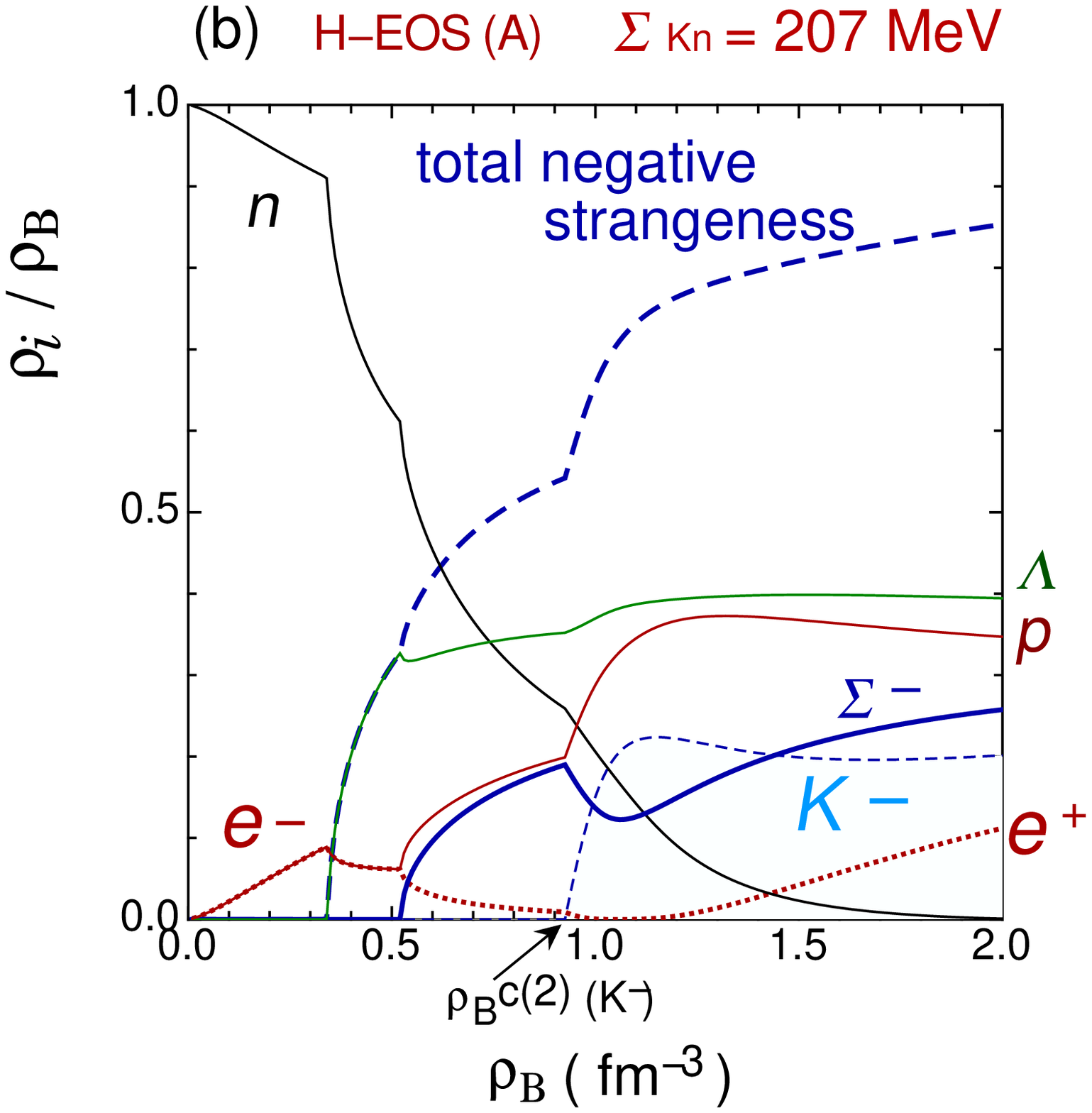}
\end{center}
\end{minipage}
\caption{(a) Particle fractions $\rho_i/\rho_{\rm B}$ in the $K^-$-condensed phase as functions of the baryon number density $\rho_{\rm B}$ for H-EOS (A) and $\Sigma_{Kn}$=305 MeV. (b) The same as in (a), but for H-EOS (A) and $\Sigma_{Kn}$=207 MeV. }
\label{fig:fraction-a}
\end{figure}

\begin{figure}[!]
\noindent\begin{minipage}[l]{0.50\textwidth}
\begin{center}
\includegraphics[height=.3\textheight]{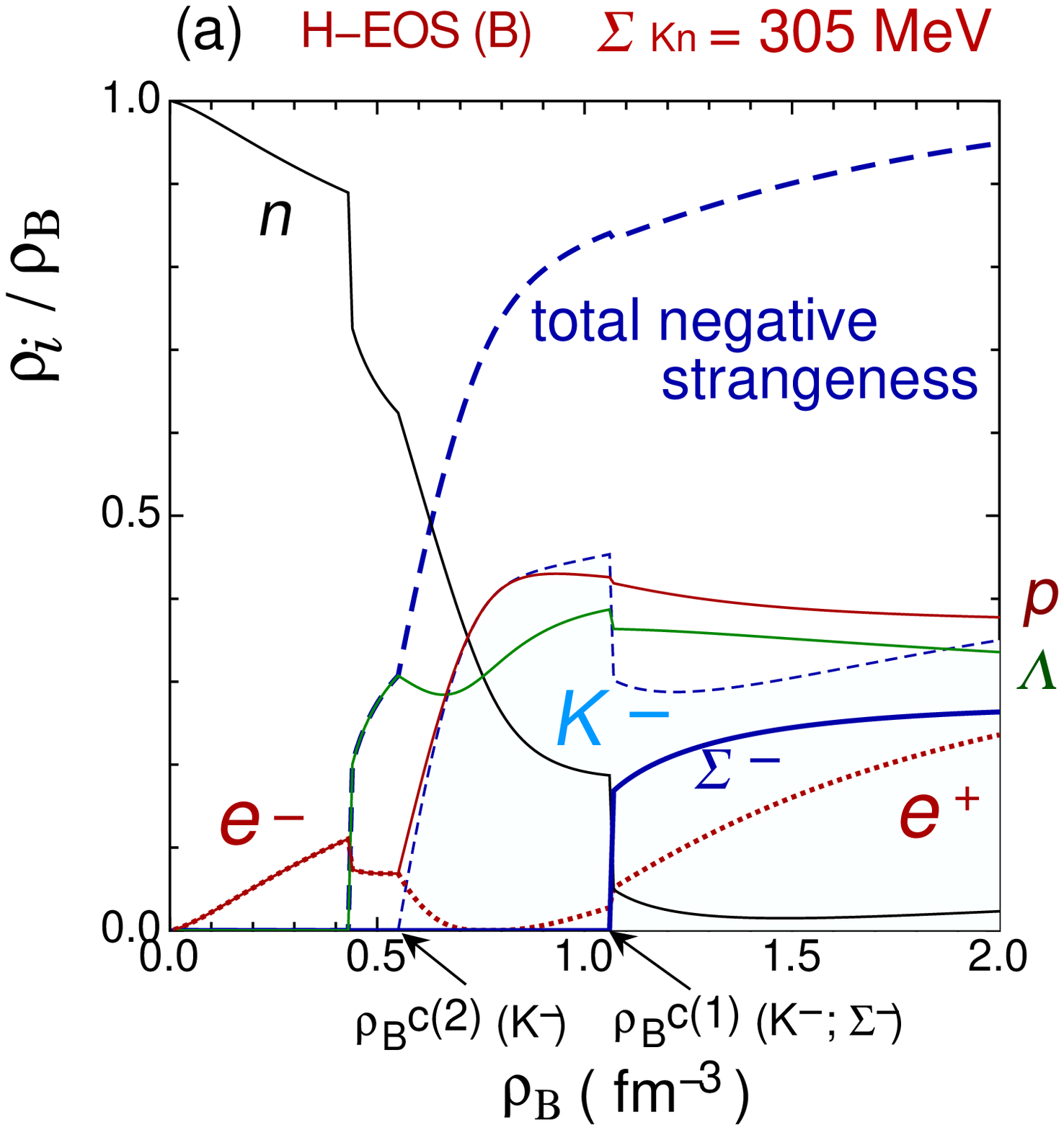}
\end{center}
\end{minipage}~
\begin{minipage}[r]{0.50\textwidth}
\begin{center}
\includegraphics[height=.3\textheight]{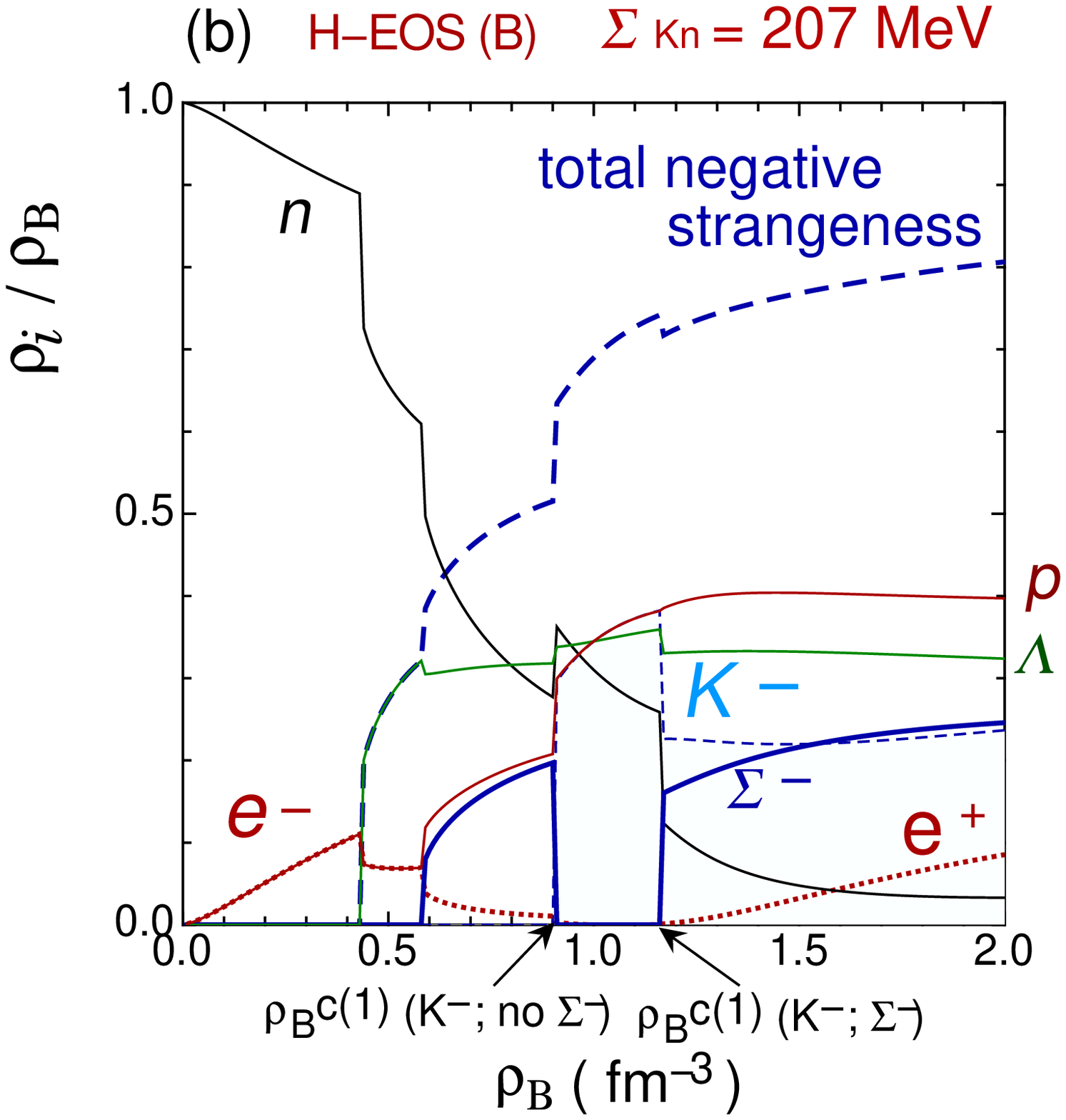}
\end{center}
\end{minipage}
\caption{(a) Particle fractions $\rho_i/\rho_{\rm B}$ in the $K^-$-condensed phase as functions of the baryon number density $\rho_{\rm B}$ for H-EOS (B) and $\Sigma_{Kn}$=305 MeV. (b) The same as in (a), but for H-EOS (B) and $\Sigma_{Kn}$=207 MeV.}
\label{fig:fraction-b}
\end{figure}

\subsection{Composition of matter in the kaon-condensed phase}
\label{subsec:composition-k}

\ \ The characteristic features of the kaon-condensed phase in hyperonic matter can be surveyed from the density dependence of the compositon of matter. In Figs.~\ref{fig:fraction-a} and \ref{fig:fraction-b}, the particle fractions $\rho_i/\rho_{\rm B}$ ($i$=$p$, $\Lambda$, $n$, $\Sigma^-$, $K^-$, $e^-$) are shown as functions of the baryon number density $\rho_{\rm B}$. Figures~\ref{fig:fraction-a}~(a) and (b) are for H-EOS(A) with $\Sigma_{Kn}$ = 305 MeV and $\Sigma_{Kn}$ = 207 MeV, respectively, and Fig.~\ref{fig:fraction-b}~(a) and (b) are for H-EOS(B) with $\Sigma_{Kn}$ = 305 MeV and $\Sigma_{Kn}$ = 207 MeV, respectively.
The long dashed lines stand for the ratio of the total negative strangeness number density $\rho_{\rm strange}$ to the baryon number density $\rho_{\rm B}$ with
\begin{equation}
\rho_{\rm strange}=\rho_{K^-}+\rho_\Lambda+\rho_{\Sigma^-} \ .
\label{eq:rho-strange}
\end{equation}
The contribution from the classical kaon part, $\rho_{K^-}/\rho_{\rm B}$, is shown by the short dashed line in each figure. 
For each curve in Figs.~\ref{fig:fraction-a} and \ref{fig:fraction-b}, only the quantity corresponding to the lowest energy minimum state in the ($\theta$, $\rho_{\Sigma^-}/\rho_{\rm B}$) plane is shown as a function of density, so that there are gaps in the quantities at the transition densities $\rho_{\rm B}^{c(1)}(K^-; {\rm no} \ \Sigma^-)$ and $\rho_{\rm B}^{c(1)}(K^-;\  \Sigma^-)$. 
One can see competitive effects between kaon condensates and $\Sigma^-$ hyperons: As the former develops around the density  
$\rho_{\rm B}^{c(1)}(K^-; {\rm no} \ \Sigma^-)$, the latter is suppressed, while as the latter develops in the kaon-condensed phase around the density $\rho_{\rm B}^{c(1)}(K^-;\  \Sigma^-)$, the former is suppressed. 

Appearance of both kaon condensates and the $\Sigma^-$ leads to considerable suppression of the electron fraction, since the negative charge of the electron is replaced by 
that of kaon condensates and $\Sigma^-$ hyperons. 
Accordingly, the charge chemical potential $\mu$ decreases with increase in the baryon number density through the relation $\rho_e=\mu^3/(3\pi^2)$. It becomes even negative above the density $\rho_{\rm B}$ = 0.77 fm$^{-3}$ for $\Sigma_{Kn}$ = 305 MeV and $\rho_{\rm B}$ = 1.06 fm$^{-3}$ for $\Sigma_{Kn}$ = 207 MeV in both the cases of H-EOS (A) and H-EOS (B). For $\mu < 0$, the positrons ($e^+$) appear in place of the electrons.

At high densities, the protons, $\Lambda$, and $\Sigma^-$ hyperons are equally populated in the kaon-condensed phase, i. e., $\rho_p/\rho_{\rm B}$, $\rho_\Lambda/\rho_{\rm B}$, $\rho_{\Sigma^-}/\rho_{\rm B}$ = 30$-$40 $\%$, whereas the neutrons almost disappear. The total negative  strangeness ratio, $\rho_{\rm strange}/\rho_{\rm B}$, gets larger with increase in density: It reaches almost unity for $\Sigma_{Kn}$ = 305 MeV and 0.8$-$0.9 for $\Sigma_{Kn}$ = 207 MeV at high densities for both H-EOS (A) and H-EOS (B). Such a high strangeness fraction implies a close connection between the kaon-condensed phase in hyperonic matter and strange matter  where $u$, $d$ and $s$ quarks are almost equally populated in quark matter. 
 
\section{Summary and Concluding Remarks}
\label{sec:summary} 

\ \ We have studied the $s$-wave kaon condensation realized in hyperonic matter based on chiral symmetry for the kaon-baryon interactions and taking into account the parameterized effective interactions between baryons. We have concentrated on interrelations between kaon condensates and negatively charged hyperons ($\Sigma^-$) and reexamined the validity of the assumption of the continuous phase transition from the noncondensed hyperonic matter to the $K^-$-condensed phase. We have also discussed the EOS and the characteristic features of the system for the fully developed kaon-condensed phase. 

The validity of the continuous phase transition for kaon condensation in hyperonic matter is summarized as follows : In cases where the condition $\omega$ = $\mu$ [Eq.~(\ref{eq:onset}) ] is satisfied at $\rho_{\rm B}$=$\rho_{\rm B}^{c(2)}(K^-)$ in the presence of the $\Sigma^-$ hyperons, there exist, in general, two energy minima in the ($\theta$, $\rho_{\Sigma^-}/\rho_B$) plane at some density intervals near the density $\rho_{\rm B}^{c(2)}(K^-)$. One is the noncondensed state with the $\Sigma^-$-mixing (P'), and the other is the $K^-$-condensed state without the $\Sigma^-$-mixing (Q'). 
If the density $\rho_{\rm B}^{c(2)}(K^-)$ is located near the onset density of the $\Sigma^-$, 
$\rho_{\rm B}^c(\Sigma^-)$, 
the state P' is a local minimum or a saddle point, and the state Q' is the absolute minimum at $\rho_{\rm B}$ = $\rho_{\rm B}^{c(2)}(K^-)$. In this case, the assumption of the continuous phase transition is not valid : 
Below $\rho_{\rm B}^{c(2)}(K^-)$, there exists a typical density $\rho_{\rm B}^{c(1)}(K^-; {\rm no} \ \Sigma^-)$ at which the energies of the two minima become equal.  Above the density $\rho_{\rm B}^{c(1)}(K^-; {\rm no} \ \Sigma^-)$, there is a discontinuous transition from the state P' to the state Q'. 
On the other hand, if the density $\rho_{\rm B}^{c(2)}(K^-)$ is located high enough from the density $\rho_{\rm B}^c(\Sigma^-)$, the state P' is always an absolute minimum. In this case, the assumption of the continuous phase transition holds true, and the onset density of kaon condensation is given by $\rho_{\rm B}^{c(2)}(K^-)$, above which kaon condensates develop continuously with increase in the baryon number density. 

In cases where the condition $\omega$ = $\mu$ is satisfied in the absence of the $\Sigma^-$ hyperon, there exists a unique minimum of the noncondensed state (P') at a point (0,0) in the ($\theta$,   $\rho_{\Sigma^-}/\rho_B$) plane, and the assumption of the continuous phase transition is kept valid. The onset density is given by $\rho_{\rm B}^{c(2)}(K^-)$. 

The above consequences on the validity of the continuous phase transition are expected to be general and should also be applied to cases where the other negatively charged hyperons such as the cascade $\Xi^-$ are present in the noncondensed ground state and where both the $s$-wave and $p$-wave kaon-baryon interactions are taken into account\cite{m06}. 

In the fully developed phase with kaon condensates, there exist two energy minima with and without the $\Sigma^-$-mixing in the ($\theta$, $\rho_{\Sigma^-}/\rho_{\rm B}$) plane at some density intervals.  At higher densities, the ground state is transferred discontinuously from the kaon-condensed state without the $\Sigma^-$-mixing to that with the $\Sigma^-$-mixing, except for the case of H-EOS (A) with the weaker $s$-wave kaon-baryon attractive interaction. The EOS of the kaon-condensed phase becomes considerably soft, since both the kaon-baryon attractions and mixing of hyperons work to lower the energy of the system. At higher densities, the stiffness of the EOS is recovered due to the increase in the repulsive interaction between baryons. As a result, in the case of the stronger $s$-wave kaon-baryon attractive interaction ($\Sigma_{Kn}$=305 MeV),
there appears a local energy minimum as a density isomer state, which suggests the existence of self-bound objects with kaon condensates on any scale from an atomic nucleus to a neutron star. 
Recently, deeply bound kaonic nuclear states have been proposed theoretically, and much discussion has been made  about the experimental achievements of them\cite{ay02,d04,yda04,k99,i01,ki03,s04,a05,mfg05,y05,ot06}. In particular, the double and/or multiple kaon clusters advocated in the recent experimental proposal by way of invariant mass spectroscopy\cite{yda04} may have a close connection with our results of kaon-condensed self-bound objects. These experimental searches for deeply bound kaonic nuclear states may provide us with important information on the existence of kaon condensation in high-density matter. 

In this paper, both kinematics and interactions associated with baryons are treated nonrelativistically. For more quantitative consideration, one needs a relativistic framework. Specifically, the $s$-wave kaon-baryon scalar attraction, which is proportional to the scalar densities for baryons in the relativistic framework, is suppressed at high densities due to saturation of the scalar densities\cite{fmmt96}. This effect is expected to make  the EOS stiffer at high densities. 

The kaon-condensed phase is important for understanding the high-density QCD phase diagram from a hadronic picture which we have taken over the relevant baryon densities. 
At high densities, however, quark degrees of freedom may appear explicitly. 
It has been shown in this paper that the kaon-condensed phase in hyperonic matter leads to large (negative) strangeness fraction, $\rho_{\rm strange}/\rho_{\rm B}\sim 1$. 
This result suggests that kaon-condensed phase in the hadronic picture may be considered as a pathway to strange quark matter. 
In a quark picture, a variety of deconfined quark phases including color superconductivity have been elaborated\cite{nhhk04}. In particular, kaonic modes may be condensed in the color-flavor locked phase\cite{bs02,kr02,b05,f05}. It is interesting to clarify the relationship between kaon condensation in the hadronic phase and that in the quark phase and a possible transition between the two phases. 

\section*{Acknowledgments}
\ \ The author is grateful to T.~Tatsumi, T.~Takatsuka, T.~Kunihiro, and M.~Sugawara for valuable discussions. He also thanks the Yukawa Institute for Theoretical Physics at Kyoto University, where this work was completed during the YKIS 2006 on ``New Frontiers on QCD''. This work is supported in part by the Grant-in-Aid for Scientific Research Fund (C) of the Ministry of Education, Science, Sports, and Culture (No. 18540288), and by the funds provided by Chiba Institute of Technology.  

\appendix
\section{Onset conditions of $\Lambda$ and $\Sigma^-$ hyperons in ordinary neutron-star matter}
\label{sec:app}

\ \ We recapitulate how hyperons $\Lambda$ and $\Sigma^-$ appear in the ordinary neutron-star matter composed of protons, neutrons and leptons within the baryon-baryon interaction models H-EOS (A) and (B). In particular, we compare onset mechanisms of the hyperon-mixing between H-EOS (A) and (B). 

\subsection{$\Lambda$-mixing}
\label{subsec:lambda}

\ \ The condition of the $\Lambda$-mixing in the noncondensed neutron-star matter is given by 
Eq.~(\ref{eq:wequil2}) with $\theta$ = 0 : 
$\mu_\Lambda=\mu_n $ 
with
\begin{subequations}\label{eq:cepln}
\begin{eqnarray}
\mu_\Lambda&=&\frac{(3\pi^2\rho_\Lambda)^{2/3}}{2M_N}
+\delta M_{\Lambda N}+V_\Lambda \ , \label{eq:cepln1} \\
\mu_n&=&\frac{(3\pi^2\rho_n)^{2/3}}{2M_N}+V_n \ ,\label{eq:cepln2}
\end{eqnarray}
\end{subequations}
together with $\mu_n=\mu_p+\mu$ [Eq.~(\ref{eq:wequil1}) with $\theta$ = 0] and the charge neutrality condition, $\rho_p=\rho_e$. 

\begin{figure}[!]
\noindent\begin{minipage}[l]{0.50\textwidth}
\begin{center}
\includegraphics[height=.30\textheight]{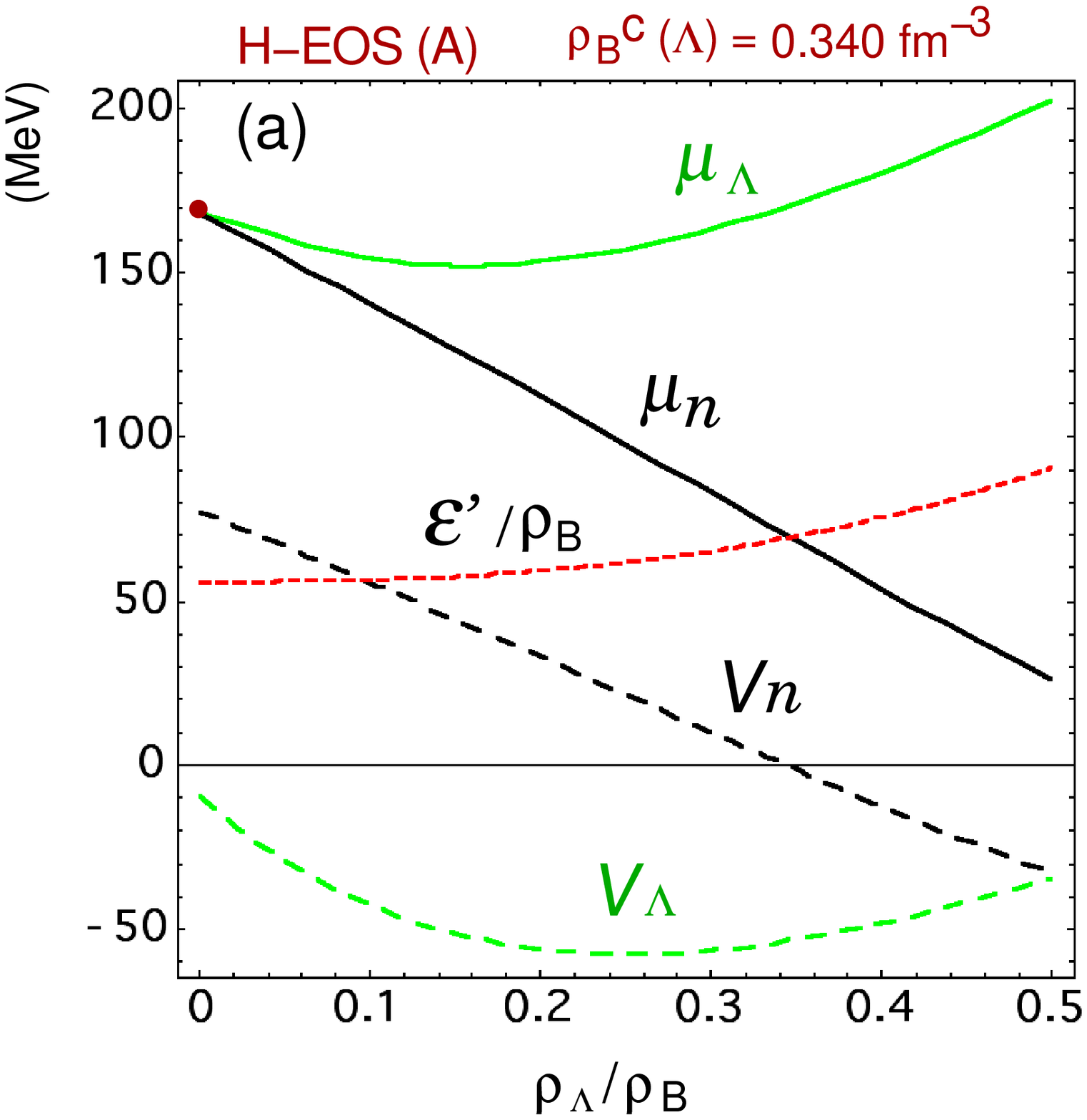}
\end{center}
\end{minipage}~
\begin{minipage}[r]{0.50\textwidth}
\begin{center}
\includegraphics[height=.3\textheight]{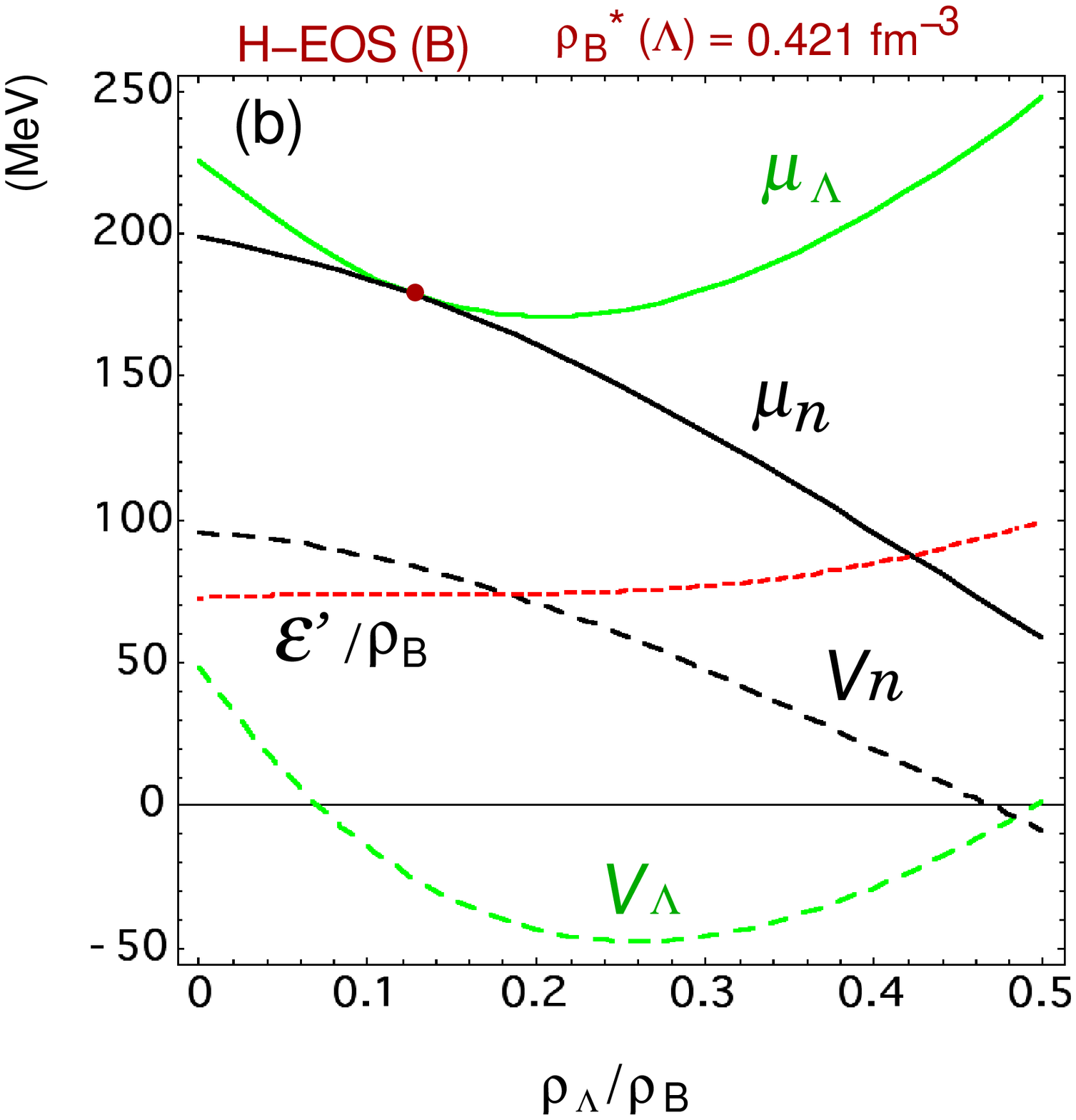}
\end{center}
\end{minipage}
\caption{Dependence of $V_{\Lambda}$, $V_n$ (dashed lines), $\mu_\Lambda$, $\mu_n$ (solid lines), and the total energy per baryon ${\cal E}'/\rho_{\rm B}$ (short dashed line) on the $\Lambda$-mixing ratio $\rho_\Lambda/\rho_{\rm B}$ in the noncondensed neutron-star matter at the minimal baryon number density 
$\rho_{\rm B}^\ast (\Lambda)$ above which the $\Lambda$-mixing condition, $\mu_\Lambda$ = $\mu_n$, is satisfied. (a) is for H-EOS (A), and (b) is for H-EOS (B). See the text for details.}
\label{fig:ceplam}
\end{figure}

In Fig.~\ref{fig:ceplam}, the baryon potentials $V_{\Lambda}$, $V_n$ (dashed lines), the chemical potentials $\mu_\Lambda$, $\mu_n$ (solid lines), and the total energy per baryon ${\cal E}'/\rho_{\rm B}$ (short dashed line) are shown as functions of the $\Lambda$-mixing ratio $\rho_\Lambda/\rho_{\rm B}$ at the minimal baryon number density 
$\rho_{\rm B}^\ast (\Lambda)$ above which the $\Lambda$-mixing condition, $\mu_\Lambda$ = $\mu_n$, is satisfied. Fig.~\ref{fig:ceplam}(a) is for H-EOS (A), and Fig.~\ref{fig:ceplam}(b) is for H-EOS (B). 
For H-EOS (A), the condition $\mu_\Lambda$=$\mu_n$ is met at $\rho_\Lambda/\rho_{\rm B}$=0, where the total energy per baryon ${\cal E}'/\rho_{\rm B}$ is a minimum. Therefore, the $\Lambda$ hyperon starts to be mixed continuously in the ground state of ($n, p, e^-$) matter at this density $\rho_{\rm B}^\ast (\Lambda)$ (= 0.340 fm$^{-3}$), which gives the onset density $\rho_{\rm B}^c (\Lambda)$.  For H-EOS (B), however, the condition $\mu_\Lambda$=$\mu_n$ is met at a nonzero value of $\rho_\Lambda/\rho_{\rm B}$ (=0.126), at which the total energy per baryon ${\cal E}'/\rho_{\rm B}$ is a local minimum, and the absolute energy minimum still lies at $\rho_\Lambda/\rho_{\rm B}$ = 0 [Fig.~\ref{fig:ceplam}(b)]. In this case, the $\Lambda$-mixing starts at a slightly higher density ($\sim$ 0.44 fm$^{-3}$) than $\rho_{\rm B}^\ast(\Lambda)$ (=0.421 fm$^{-3}$) with a nonzero value of $\rho_{\Lambda}/\rho_{\rm B}$.

In the cases of both H-EOS (A) and (B), the dependence of the chemical potentials $\mu_\Lambda$ and $\mu_n$ on the $\Lambda$-mixing ratio is closely correlated with that of the baryon potentials $V_\Lambda$ and $V_n$, respectively, as seen in Figs.~\ref{fig:ceplam}~(a) and (b). In fact, the difference of the onset mechanisms of the $\Lambda$-mixing between the cases H-EOS (A) and (B) stems from the difference of the dependence of the baryon potentials on the $\Lambda$-mixing ratio between H-EOS (A) and (B).

\subsection{$\Sigma^-$-mixing}
\label{subsec:sigma}

\ \ The condition of the $\Sigma^-$-mixing in the noncondensed ($n$, $p$, $\Lambda$, $e^-$) matter is given by 
Eq.~(\ref{eq:wequil3}) with $\theta$ = 0 : $ \mu_{\Sigma^-}=\mu_n + \mu$ 
with
\begin{equation}
\mu_{\Sigma^-}=\frac{(3\pi^2\rho_{\Sigma^-})^{2/3}}{2M_N}
+\delta M_{\Sigma^- N}+V_{\Sigma^-}  
\label{eq:cepsign}
\end{equation}
 and (\ref{eq:cepln2}), together with $\mu_n=\mu_p+\mu$ [Eq.~(\ref{eq:wequil1}) with $\theta$ = 0], $\mu_\Lambda$ = $\mu_n$ [Eq.~(\ref{eq:wequil2}) with $\theta$ = 0], and the charge neutrality condition, $\rho_p=\rho_e$. 

\begin{figure}[!]
\noindent\begin{minipage}[l]{0.50\textwidth}
\begin{center}
\includegraphics[height=.3\textheight]{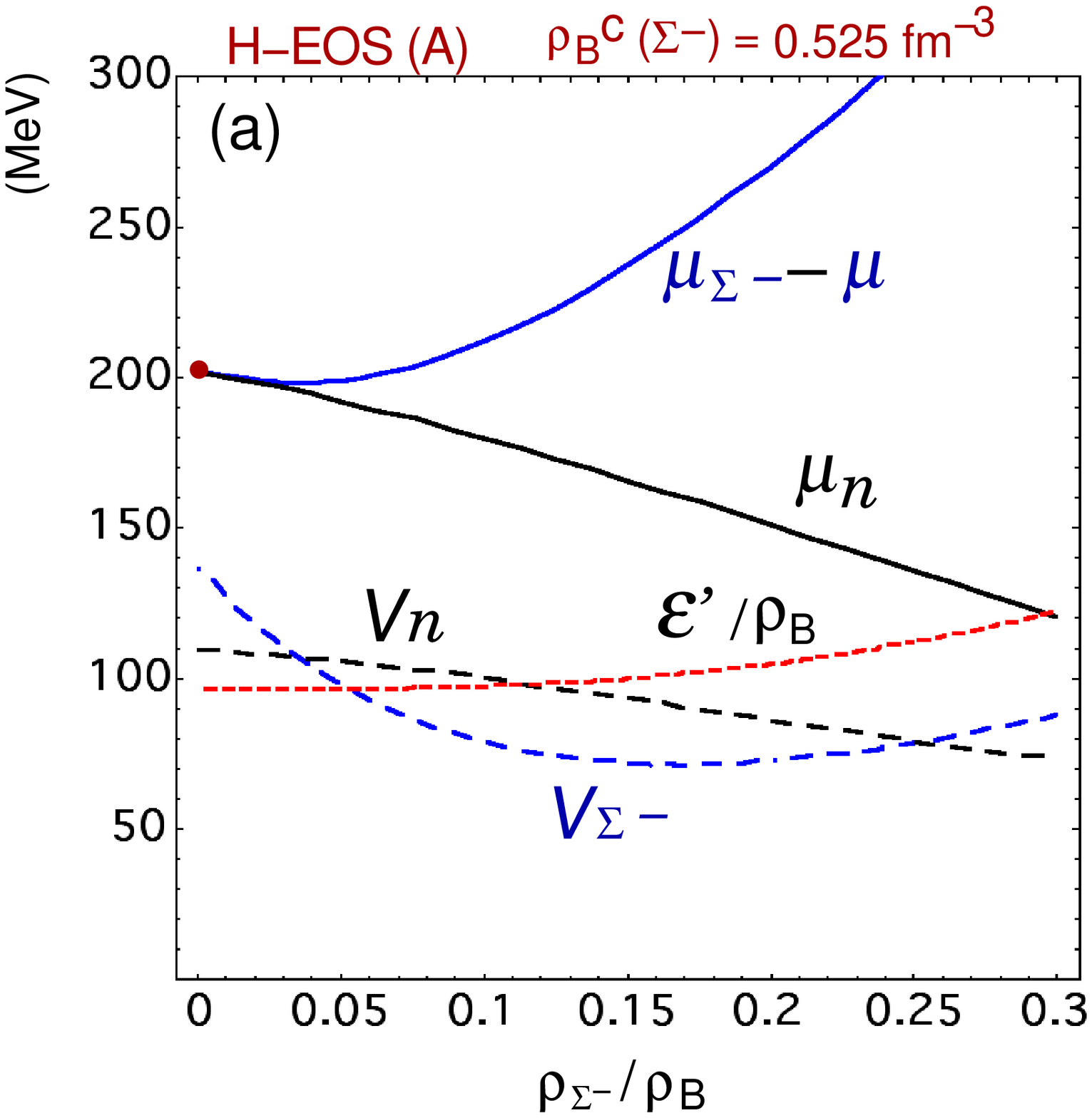}
\end{center}
\end{minipage}~
\begin{minipage}[r]{0.50\textwidth}
\begin{center}
\includegraphics[height=.3\textheight]{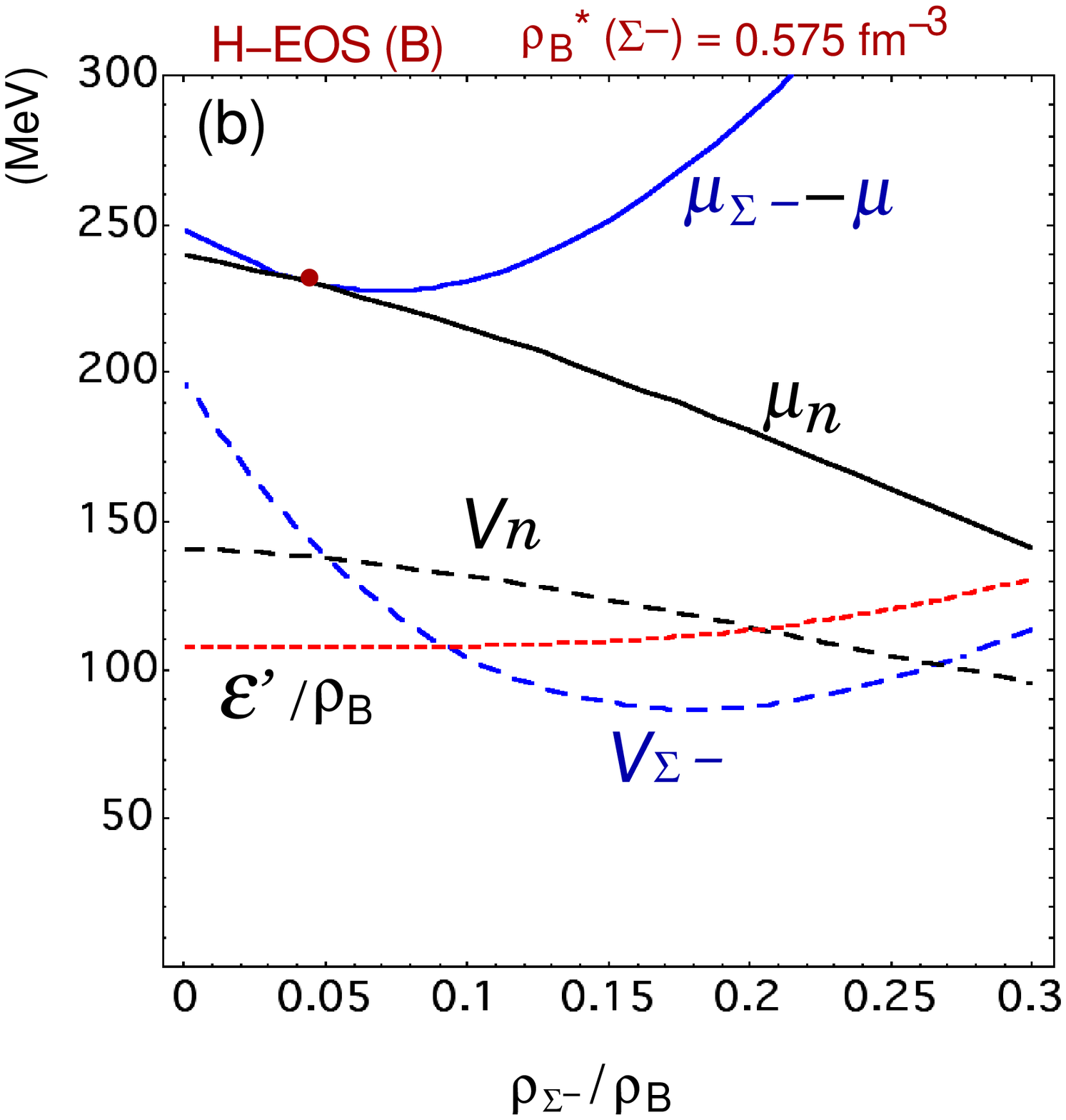}
\end{center}
\end{minipage}
\caption{Dependence of the baryon potentials $V_{\Sigma^-}$, $V_n$ (dashed lines), the difference of the $\Sigma^-$ chemical potential and charge chemical potential, $\mu_{\Sigma^-}-\mu$ (solid line), the neutron chemical potential $\mu_n$ (solid line), and the total energy per baryon ${\cal E}'/\rho_{\rm B}$ (short dashed line) on the $\Sigma^-$-mixing ratio $\rho_{\Sigma^-}/\rho_{\rm B}$ in the noncondensed ($n$, $p$, $\Lambda$, $e^-$) matter at the minimal baryon number density 
$\rho_{\rm B}^\ast (\Sigma^-)$ above which the $\Sigma^-$-mixing condition, $\mu_{\Sigma^-}=\mu_n + \mu$, is satisfied. (a) is for H-EOS (A), and (b) for H-EOS (B). See the text for details.}
\label{fig:ceps}
\end{figure}

In Fig.~\ref{fig:ceps}, we depict the baryon potentials $V_{\Sigma^-}$, $V_n$ (dashed lines), the difference of the $\Sigma^-$ chemical potential and charge chemical potential, $\mu_{\Sigma^-}-\mu$ (solid line), the neutron chemical potential $\mu_n$ (solid line), and the total energy per baryon ${\cal E}'/\rho_{\rm B}$ (short dashed line) as functions of the $\Sigma^-$-mixing ratio $\rho_{\Sigma^-}/\rho_{\rm B}$ in the noncondensed ($n$, $p$, $\Lambda$, $e^-$) matter at the minimal baryon number density 
$\rho_{\rm B}^\ast (\Sigma^-)$ above which the $\Sigma^-$-mixing condition, $\mu_{\Sigma^-}=\mu_n + \mu$, is satisfied. (a) is for H-EOS (A), and (b) is for H-EOS (B). One finds the results on the onset mechanism of the $\Sigma^-$-mixing similar to the case of the $\Lambda$-mixing : For H-EOS (A) [Fig.~\ref{fig:ceps}~(a)], the onset density $\rho_{\rm B}^c(\Sigma^-)$ is given by $\rho_{\rm B}^c(\Sigma^-)$=$\rho_{\rm B}^\ast(\Sigma^-)$ (= 0.525 fm$^{-3}$) with $\rho_{\Sigma^-}/\rho_{\rm B}$ = 0, and the $\Sigma^-$-mixing ratio increases continuously from zero with increase in the baryon number density. For H-EOS (B) [Fig.~\ref{fig:ceps}~(b)], the onset density $\rho_{\rm B}^c(\Sigma^-)$ ($\sim$ 0.59 fm$^{-3}$) is slightly larger than the minimal density $\rho_{\rm B}^\ast (\Sigma^-)$ (=0.575 fm$^{-3}$)  satisfying the $\Sigma^-$-mixing conditions,
and the $\Sigma^-$-mixing starts with a nonzero value of $\rho_{\Sigma^-}/\rho_{\rm B}$. 
The dependence of the chemical potentials $\mu_{\Sigma^-}$ and $\mu_n$ on the $\Sigma^-$-mixing ratio is correlated with that of the baryon potentials $V_{\Sigma^-}$ and $V_n$, respectively, which leads to the difference of the onset machanisms of the $\Sigma^-$-mixing between H-EOS (A) and (B).

\end{document}